\documentclass{emulateapj}

\usepackage{psfig,natbib,amsmath}   

\newcommand{\mic}{\,$\mu$m }
\newcommand{\micpa}{\,$\mu$m}          
\newcommand{\muJy}{\,$\mu$Jy }
\newcommand{\muJypa}{\,$\mu$Jy}

\bibpunct[]{(}{)}{;}{a}{}{,}

\shorttitle{The evolution of Infrared Galaxies at  0\,$\ltapp$\,$z$\,$\ltapp$\,1}
\shortauthors{E.\,Le Floc'h et al.}

\begin{document}
\def\gtapp
{\mathrel{\hbox{\raise0.3ex\hbox{$>$}\kern-0.8em\lower0.8ex\hbox{$\sim$}}}}
\def\ltapp
{\mathrel{\hbox{\raise0.3ex\hbox{$<$}\kern-0.75em\lower0.8ex\hbox{$\sim$}}}}
\def\ts{\thinspace}


\title{Infrared luminosity functions from the Chandra Deep Field South :
the  {\it Spitzer\,} view on the 
history of dusty star formation at 0\,$\ltapp$\,$z$\,$\ltapp$\,1
$^{1}$}

\altaffiltext{1}{Based on observations made with {\it Spitzer}, 
 operated by the Jet Propulsion Laboratory under NASA
contract 1407.}

\slugcomment{Accepted for publication in The Astrophysical Journal,
June 16th, 2005}


\author{Emeric~Le~Floc'h$^{\rm a}$, 
 Casey~Papovich$^{\rm a}$,
Herv\'e~Dole$^{\rm b}$,
Eric\,F.~Bell$^{\rm c}$, 
Guilaine~Lagache$^{\rm b}$,
George\,H.~Rieke$^{\rm a}$,
Eiichi~Egami$^{\rm a}$,
Pablo\,G.~P\'erez-Gonz\'alez$^{\rm a}$,
Almudena~Alonso-Herrero$^{\rm d}$, 
Marcia\,J.~Rieke$^{\rm a}$,
Myra~Blaylock$^{\rm a}$, 
Charles\,W.~Engelbracht$^{\rm a}$,
Karl\,D.~Gordon$^{\rm a}$,
Dean\,C.~Hines$^{\rm a,e}$,
Karl\,A.~Misselt$^{\rm a}$, 
Jane\,E.~Morrison$^{\rm a}$
and Jeremy~Mould$^{\rm f}$ }
\affil{$^{\rm a}$ Steward Observatory,
University of Arizona, Tucson, AZ 85721, USA \\ 
$^{\rm b}$ Institut d'Astrophysique
Spatiale, Universit\'e Paris Sud, F-91405 Orsay Cedex, France \\
$^{\rm c}$ Max-Planck-Institut f\"{u}r Astronomie, K\"{o}nigstuhl 17, D-69117 
Heidelberg, Germany \\
$^{\rm d}$ Instituto de Estructura de la Materia, CSIC, E-28006, Madrid, Spain \\
 $^{\rm e}$Space Science Institute, 4750 Walnut Street,
Suite 205 Boulder, Colorado 80301, USA \\
 $^{\rm f}$ National Optical Astronomy Observatory, P.O.\,Box~26732, Tucson, AZ 85726, USA
}

\begin{abstract} 

We analyze a sample of $\sim$\,2\,600 MIPS/{\it Spitzer\,} 24\mic
sources brighter than $\sim$\,80\muJy and located in the Chandra Deep
Field South to characterize the evolution of the comoving infrared
(IR) energy density of the Universe up to $z$\,$\sim$\,1.  Using
published ancillary optical data we first obtain a nearly complete
redshift determination for the 24\mic objects associated with
$R$\,$\ltapp$\,24\,mag counterparts at $z$\,$\ltapp$\,1.  These
sources represent $\sim$\,55-60\% of the total MIPS 24\mic population
with $f_{24\mu m}$\,$\gtapp$\,80\muJypa, the rest of the sample likely
lying at higher redshifts.  We then determine an estimate of their
total IR luminosities using various libraries of IR spectral energy
distributions.  We find that the 24\mic population at
0.5\,$\ltapp$\,$z$\,$\ltapp$\,1 is dominated by ``Luminous Infrared
Galaxies'' (i.e., 10$^{11}$\,L$_{\odot}$\,$\leq$\,L$_{\rm
  IR}$\,$\leq$\,10$^{12}$\,L$_{\odot}$), the counterparts of which
appear to be also luminous at optical wavelengths and tend to be more
massive than the majority of optically-selected galaxies.  A
significant number of fainter sources
(5$\times$10$^{10}$\,L$_{\odot}$\,$\ltapp$\,L$_{\rm
  IR}$\,$\leq$\,10$^{11}$\,L$_{\odot}$) are also detected at similar
distances.  We finally derive 15\mic and total IR luminosity functions
(LFs) up to $z$\,$\sim$\,1. In agreement with the previous results
from $ISO$ and SCUBA and as expected from the MIPS source number
counts, we find very strong evolution of the contribution of the
IR-selected population with lookback time.  Pure evolution in density
is firmly excluded by the data, but we find considerable degeneracy
between strict evolution in luminosity and a combination of increases
in both density and luminosity ($L^\star_{\rm
  IR}$\,$\propto$\,(1+$z$)$^{3.2_{-0.2}^{+0.7}}$, $\phi^\star_{\rm
  IR}$\,$\propto$\,(1+$z$)$^{0.7_{-0.6}^{+0.2}}$).  A significant
steepening of the faint end slope of the IR luminosity function is
also unlikely, as it would overproduce the faint 24\mic source number
counts.  Our results imply that the comoving IR energy density of the
Universe evolves as (1+$z$)$^{3.9\pm0.4}$ up to $z$\,$\sim$\,1 and
that galaxies luminous in the infrared (i.e., L$_{\rm
  IR}$\,$\geq$\,10$^{11}$\,L$_\odot$) are responsible for 70$\pm$15\%
of this energy density at $z$\,$\sim$\,1. Taking into account the
contribution of the UV luminosity evolving as (1+$z$)$^{\sim2.5}$, we
infer that these IR-luminous sources dominate the star-forming
activity beyond $z$\,$\sim$\,0.7.  The uncertainties affecting these
conclusions are largely dominated by the errors in the $k$-corrections
used to convert 24\mic fluxes into luminosities.

\end{abstract}

\keywords{ galaxies: high-redshift ---  infrared: galaxies ---
 cosmology: observations}

\section{Introduction}

The successful launch of the {\it Spitzer Space Telescope\,}
\citep{Werner04} recently opened a new exciting window on the deep
infrared (IR) Universe. {\it Spitzer\,} operates between 3.6 and
160\mic with unprecedented sensitivity and better spatial resolution
compared to previous infrared satellites (e.g., {\it IRAS,\,} {\it
  ISO}).  Directly probing the dust emission or the redshifted
signature of distant stellar populations, its first extragalactic
surveys have already unveiled a huge number of faint and high redshift
sources \citep[see for instance][ in the {\it ``Spitzer\,} Special
  Edition -- volume\,154'' of the {\it ApJ
    Supplement}]{Fazio04,Eisenhardt04,Papovich04,Marleau04,Lonsdale04,Chary04,Yan04,Dole04}.
{\it Spitzer} therefore provides new opportunities to determine the IR
properties of galaxies in the general context of cosmic evolution.

High redshift sources detected by {\it Spitzer\,} in the mid- and
 far-infrared wavelength range (i.e.,
 8\mic$\,\ltapp$\,$\lambda$\,$\ltapp$\,1000\micpa) are characterized
 by intrinsically very high luminosities
 \citep{Egami04a,Frayer04,Ivison04,LeFloch04}.
They appear as the distant analogs of the local Luminous and Ultra-Luminous
InfraRed Galaxies (respectively LIRGs: $10^{11}$\,L$_{\odot} \leq$
L$_{\rm IR} = $L$[8-1000\mu m] \leq 10^{12}$\,L$_{\odot}$, and ULIRGs:
L$_{\rm IR} \geq 10^{12}$\,L$_{\odot}$, see the review by
\citealt{Sanders96}).  Such infrared-luminous sources\footnote{We
adopt in this paper the more general expression of ``infrared-luminous
galaxies'' to denote sources characterized by L$_{\rm IR} \geq
10^{11}$\,L$_{\odot}$.}  emit the bulk of their energy 
as dust-reprocessed thermal IR emission powered by
 embedded star formation or by accreted material
surrounding
supermassive black holes. They were first discovered in the
nearby Universe with ground-based observations \citep{Rieke72}. After
being systematically catalogued by {\it IRAS\,} \citep{Soifer87b},
they were found to be locally very rare and to only account for $\sim$5\% of the
total infrared energy emitted by galaxies at low redshift
\citep{Soifer91,Kim98a}. Nevertheless, there is  clear evidence that
they were significantly more numerous earlier in  cosmic history.
In the past few years, deep observations performed in the infrared by
{\it ISO} and in the submillimeter by the SCUBA camera have
revealed  strong evolution of these luminous sources with
lookback time \citep{Smail97,Blain99b,Elbaz99,Serjeant01,Dole01},
that is also apparent in the population of radio
sources at \muJy flux levels \citep[e.g.,][]{Cowie04}. 
Characterized by a high redshift space density several
orders of magnitude larger than predicted by non-evolving models,
infrared luminous galaxies contribute a significant fraction of the distant
starbursting activity and play a crucial role in the formation of
massive spheroidals throughout the cosmic ages
\citep[e.g.,][]{Flores99,Gispert00,Franceschini01,Chary01,Blain02,Chapman03}.

This strong evolution of infrared-selected sources has also been 
clearly seen by {\it Spitzer\,} (\citealt{Chary04},
\citealt{Papovich04}, hereafter P04; \citealt{Marleau04,Dole04}).
One of the most interesting results of the  {\it Spitzer\,} deep
surveys is the behavior of the differential number counts at 24\micpa.
These counts  turn over at fluxes lower than had been
expected based on ``pre-launch'' models.
\citet{Lagache04} suggest that these counts reveal even more luminous
galaxies
$z$\,$\gtapp$\,1.5 than expected, though one could argue that they
 can also originate
from a steeper faint-end slope of the infrared luminosity function at
more modest redshifts.
To better understand the nature of the sources
responsible for this turn-over 
and also more generally the role of infrared galaxies in  cosmic
evolution, we examine in this paper 
the evolution of the comoving IR energy density with redshift
up to $z$\,$\sim$\,1. This study is based on a sample of
24\micpa-selected {\it Spitzer\,} sources 
within the Chandra Deep Field South and
characterized by
 redshifts taken from  the literature.
A companion publication by Bell et al. (2005) explores in more detail
the role played by these sources in  the decline
of the star formation history since $z$\,$\sim$\,0.7. 

The paper 
 is organized as follows.
In Sect.\,2 we  describe the infrared and  optical data used
in this study, while Sect.\,3 outlines the results of the
cross-correlations that we performed among catalogs to determine
the  redshifts of infrared-selected sources. In Sect.\,4 we
study the contribution of these infrared galaxies to the total counts
at 24\mic for various redshift limits and present a comparison with the
predictions from various recent models of IR galaxy evolution.
 Using different libraries of templates published in the literature,
we further derive in Sect.\,5 an estimate of the total infrared
luminosities of these sources based on our mid-infrared observations.
In Sect.\,6 we analyze a few properties of their optical counterparts
and we finally explore in
Sect.\,7 how the evolution of the infrared luminosity function at $0
\ltapp z \ltapp 1$ in the CDFS can be constrained using our
data. Interpretations are discussed in Sect.\,8, and we give our
conclusions in Sect.\,9. Throughout this work, we assume a
$\Lambda$CDM cosmology with H$_0$\,=\,70~km~s$^{-1}$\,Mpc$^{-1}$,
$\Omega_m$\,=\,0.3 and $\Omega_{\lambda}\,=\,0.7$ \citep{Spergel03}.
Unless explicitly stated, magnitudes are quoted within the Vega
system. We also adopt the universal Initial Mass Function
from \citet{Salpeter55}.

\section{The data}

\subsection{24\mic imaging}

The region of the Chandra Deep Field South (hereafter CDFS,
$\alpha=3^h32^m00^s$, $\delta=-27^{\rm o}35'00''$, J2000) was observed
with the MIPS instrument \citep{Rieke04} on-board the {\it Spitzer
Space Telescope\,}  in January 2004 as part of the MIPS
Guaranteed Time Observing program.  These observations were performed
over a total field of $\sim$1.45$\times$0.4\,=\,0.6\,deg$^2$
with the so-called
``Slow Scan'' technique, a MIPS observing mode that allows the coverage of
large sky areas with high efficiency. The detector at 24\mic uses a
2.45\arcsec\,pixel size array of 128$\times$128 elements and the image 
at this wavelength is
characterized by a Point Spread
Function (PSF) with a Full-Width at Half Maximum (FWHM) of $\sim$6\arcsec.  
The effective integration time per sky pixel
was $\sim$1380\,s.  Data reduction was carried out using the MIPS Data
Analysis Tool \citep[DAT,][]{Gordon05}. The mosaic  finally obtained
has a pixel scale of half that of the physical detector pixel. A color version
of this 24\mic map can be seen in the publication by \citet{Rigby04}.
A sub-region of this field is also illustrated in Figure\,1.

\begin{minipage}[b]{8.5cm}
\vskip .3cm 
\centerline{\psfig{file=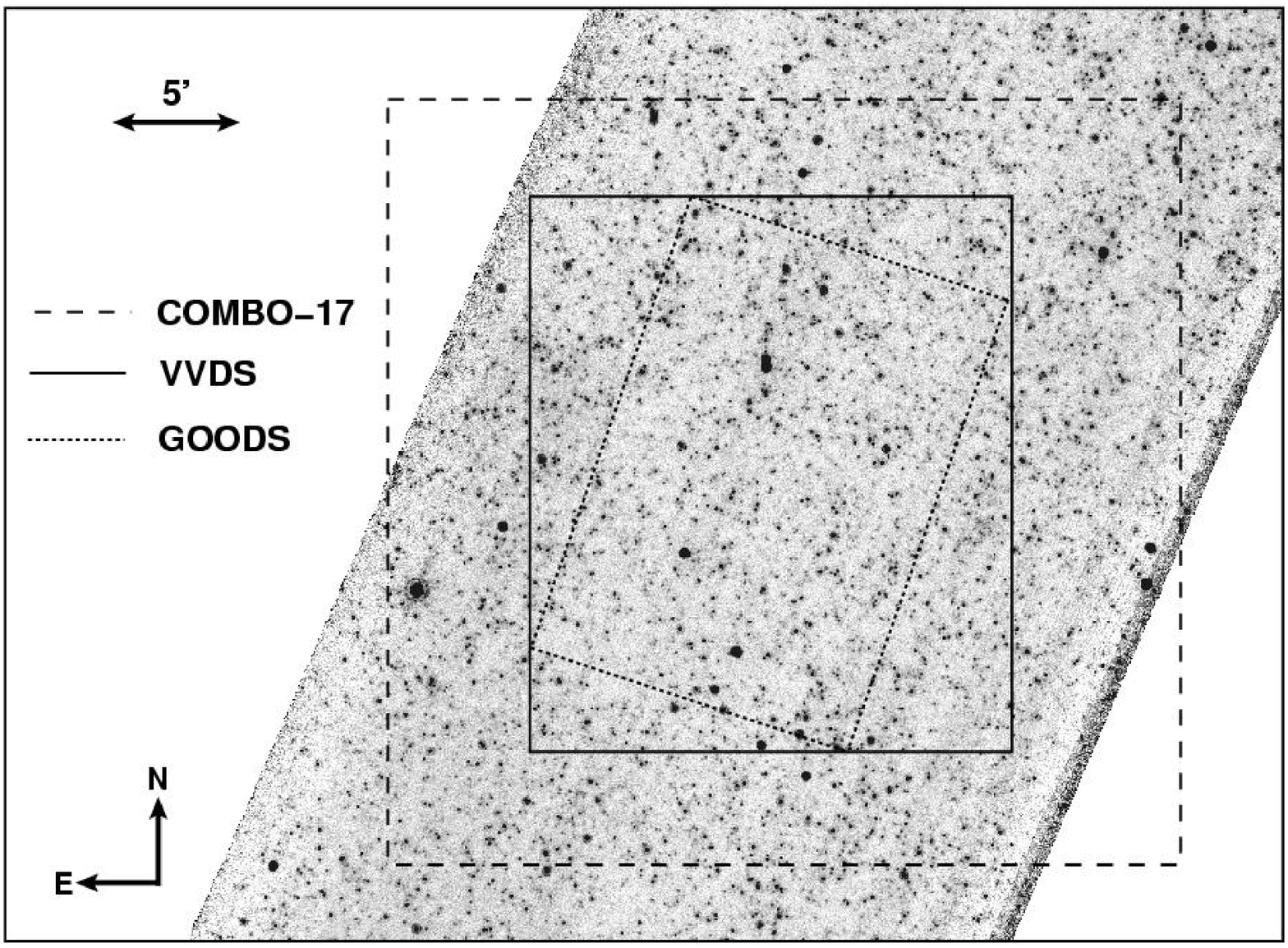,width=8.5cm,angle=0}}
\vskip 0.2cm \figcaption{ A sub-image of the CDFS 24\mic observations
obtained with MIPS, centered at $\alpha=3^h32^m28^s$ and
$\delta=-27^{\rm o}48'27''$ (J2000).  The fields of view respectively
covered by COMBO-17 (dashed line), VVDS (solid line) and GOODS (dotted
line) are also indicated (see text for details).  The 24\mic sources
considered in this paper are located within the 775\,arcmin$^2$ of
overlap between MIPS and COMBO-17.  }\vskip .2cm
\label{fig:z_distrib}
\end{minipage}

Since most sources are unresolved in our data, 
extraction and photometry were performed
using the PSF  fitting technique of the DAOPHOT software
\citep{Stetson87}. An empirical point spread function was
constructed from the brightest objects found in our mosaic, 
and it was subsequently fitted to all the sources detected in the map.
Allowing for  multiple-match fitting to deal with blended cases, we
derived the flux density of each source from the scaled fitted PSF
and finally applied a slight correction to
account for the finite size of the modeled point spread function.  We
also performed extensive simulations adding and recovering artificial
sources in the data, which allowed us to derive an 80\% completeness
limit at $\sim$83\muJypa. 
Contamination by false sources appears in our detection catalog 
at fluxes fainter than 90\muJypa. 
 In the full
  sample of objects brighter than the 80\% completeness limit, we
  estimate such a contamination 
to be around~0.6\%.  A more detailed
description of  our technique
is presented by \citet{Papovich04} who also discuss the evolution of
the completeness limit and the false source fraction
as a function of the flux at 24\mic (see their figure~1).

Distortion effects are efficiently removed by the DAT.  The final
astrometric uncertainty of the mosaics produced by the pipeline 
 therefore originates from the pointing reconstruction.
To quantify
this  uncertainty in our data, we cross-identified the
brightest sources detected at 24\mic with the {\it Two Micron All Sky
Survey\,} (2MASS) catalog \citep{Jarrett00}. A systematic offset of
$\sim$0.6\arcsec~was observed in the scan direction\footnote{This
offset seems to be due to
 systematics related to the position of the MIPS scan mirror.
It affects the headers of the data obtained before May~2004
 ({\it Spitzer\,} Science Center, priv. communication).}
and subsequently removed for our source
catalog to match the 2MASS coordinates.  We estimate that
the remaining scatter
relative to 2MASS is better than 0.3$\arcsec$~rms.

\subsection{Optical-spectroscopic redshifts}

\label{sect:spec_z}

To derive the redshifts of the MIPS 24\micpa--selected sources, we
retrieved from the literature publicly available catalogs of optical
spectroscopic surveys in the CDFS such as the ``VIMOS VLT
Deep Survey''\footnote{http://cencosw.oamp.fr/EN/index.en.html}
\citep[hereafter VVDS,][]{LeFevre04}, the ESO/FORS2 survey performed
by the GOODS Legacy
team\footnote{http://www.eso.org/science/goods/spectroscopy/products.html}
\citep{Vanzella05}, and the follow-up of X-ray sources described by
\citet{Szokoly04}\footnote{http://www.mpe.mpg.de/~mainieri/cdfs\_pub/}.
The VVDS consortium has released redshifts for 1599 sources located in
an area of 21$\times$21.6\,arcmin$^2$ with an overall redshift
measurement completeness of $\sim$88\% down to $I_{AB} \leq 24$. 
For each identification, a flag indicates the reliability
level of the corresponding measurement; 1457 sources in this
survey are classified with more than 75\%  confidence in the
redshift determination.  The GOODS catalog provides 234 redshifts
obtained within a slightly smaller area of the CDFS
(10$\times$15\,arcmin$^2$). They are also tentatively classified into
three categories depending on their reliability,
and 150 sources in this catalog are
thought to have a secure redshift measurement. Finally, \citet{Szokoly04}
present a spectroscopic follow-up of {\it Chandra} X-ray sources, with
redshifts  obtained for 168 objects. Among those, 126 are considered 
to have  unambiguous identifications.

We combined the catalogs of these three optical surveys to create a
single list of 1\,941 spectroscopic redshifts with their corresponding
flags. Because of some overlap between the different observed
regions, several redshifts were sometimes assigned to a given
single source.  In case of discrepant estimates, 
we  kept the one flagged with the
highest confidence in the redshift measurement. 

The MIPS image at 24\mic entirely covers the areas observed by these
spectroscopic surveys. The VVDS and GOODS fields of view can be
seen in Figure\,1.

\subsection{Optical-photometric redshifts}
In addition to the spectroscopic redshifts previously described, we also
made  extensive use of the photometric redshifts from the
COMBO-17 survey \citep[``{\it Classifying Objects by Medium-Band
Observations in 17 filters}'',][]{Wolf04}. COMBO-17 observed a
31.5$\times$30\,arcmin$^2$ region of the CDFS (see Fig.\,1) through a set of 5
broad-band and 12 narrow-band filters, which allowed the determination
of accurate spectral energy distributions (SEDs) and absolute magnitudes for
several thousand optically-selected sources \citep[see
also][]{Wolf03}. Each  SED was analyzed using a library
of representative templates for various spectral types, and a 
 redshift probability distribution was subsequently 
derived for each  source.
 For a
total of 24\,217~objects, 
these  distributions were successfully fitted
 with  Gaussian-like functions, the mean and the variance of which
led to well-constrained estimates of
 photometric redshifts with  corresponding uncertainties.

These
  redshifts
are accurate to 1\% in $\delta_z/(1+z)$ for galaxies with $R
\leq$~21\,mag, and they are mostly reliable (i.e., $\delta_z/(1+z)
\leq 10\%$) for all sources at
 $z$\,$\ltapp$\,1.2 and brighter than $R$\,$\sim$\,24\,mag 
(11\,422~objects). The yield of high weight
redshifts drops  steeply for $z$\,$\gtapp$\,1.2 or $R$\,$\gtapp$\,24\,mag,
so we did not consider such faint or distant objects.
Another redshift estimate is also provided
for nearly every detection of the survey (62\,366~sources) based on
the peak of the computed probability distribution. Because it is 
less reliable, it
should be used with caution \citep{Wolf04} and was therefore also not considered
in this work.  
 About $\sim$85\%
of the region covered by COMBO-17 overlaps with our MIPS 24\mic mosaic 
 and  the spectroscopic optical surveys. 
This overlapping area has a total field of view of 775\,arcmin$^2$
(see 
Fig.\,1).

\section{Optically-selected redshifts of MIPS sources}

\subsection{Cross-correlation between the infrared and the optical data}

From the full  MIPS catalog we first selected the 24\mic sources
located in the common area covered by {\it Spitzer\,} and the aforementioned
optical redshift surveys
(i.e., 775\,arcmin$^2$).
In this region of overlap we
 detected 5\,589~objects at 24\micpa, with 3\,616 of those having a
flux greater than the 80\% completeness limit of 83\muJypa.

We cross-correlated this 24\mic sub-sample with the optical data using
a tolerance radius of 2\arcsec~for matching sources. This choice was
first motivated by the rather large FWHM of the MIPS 24\mic PSF
($\sim$6\arcsec) compared to that typically seen in optical images.
As already observed in local interacting systems
\citep[e.g.,][]{LeFloch02,Charmandaris04a,Gallais04}, it also accounts
for the physical shift that could be present between the location of
the infrared emission and the brightest optical component of distant
mergers (2\arcsec~corresponds to a linear projection of $\sim$15\,kpc
on the sky at $z$\,$\sim$\,1). 
It is yet
reasonably small given the position accuracy of the source centroids
at 24\mic (better than 0.5\arcsec~rms). Taking a larger value also
 increases the risk of associating MIPS sources with wrong
optical counterparts in the case of multiple matches. When a double
match was found, 
we  selected the closest object.
 We  ignored the cases where  three or more optical
sources could be associated with a given 24\mic\,detection.

We first correlated
 the 24\mic source catalog with the list of spectroscopic redshifts.
A total of 543~matches
was found in this cross-identification, of which 465~objects
are flagged to have a high-confidence redshift measurement.  The
fraction of multiple matches was only 1\%.  At this stage, we only
kept the identifications with a secure redshift determination. Those
465~sources 
represent only 8\% of the infrared-selected sub-sample, which 
emphasizes  the critical need for  using 
photometric redshifts. We thus
cross-correlated the rest of the data with the catalog of
COMBO-17. We found 2\,170~MIPS sources (1\,987 single and 183 double
matches) identified with a photometric redshift below $z$\,=\,1.2 and
an optical counterpart brighter than $R$\,=\,24\,mag. 
For 9~objects
selected at 24\mic (less than 0.5\% of the sample) 
three
 possible matches were found
within 2\arcsec~around the MIPS source. 
They were not further considered.
In total, we assembled 
a catalog of 2\,635 MIPS sources (of which 1\,962 are
brighter than the 80\% completeness limit of the 24\mic survey)
associated either with
a reliable spectroscopic redshift or a clearly-constrained
photometric redshift. 
Virtually all are at $z$\,$\ltapp$\,1.2, since the yield of
 values at higher redshift is very low with both 
spectroscopy (due to the ``redshift desert'') and 
COMBO-17.
We also identified another set of 1\,681
MIPS sources with an optical counterpart in the COMBO-17
catalog but without any   reliable redshift (271 of them have
$f_{24\mu m}$\,$\geq$\,83\muJy and are
brighter than $R$\,=\,24\,mag).

\subsection{Redshift uncertainties}

Since only a very small fraction of the MIPS sources have been
identified with secure spectroscopic redshifts, it is worth
looking at  the typical uncertainties of
 the other {\it photometric\,} redshifts 
characterizing the 24\micpa-selected sources. The accuracy of 
the COMBO-17 classification  decreases  for sources fainter
than $R$\,$\sim$\,22\,mag. This may have non-negligible effects when estimating
e.g., source densities as a function of lookback time, especially
when  the uncertainties 
become comparable to the redshift bins in which galaxy
properties are averaged. Given that the 24\mic sample is by definition selected through the
emission by warm dust, one may furthermore question whether the implied
extinction at optical wavelengths can lead to a more significant
redshift mis-classification in the specific case of  the most
luminous (i.e., dust-obscured) MIPS sources.

In Figures\,2a \&~2b we compare high-confidence spectroscopic redshifts
of optically-selected field galaxies
and  MIPS 24\mic detections
with their photometric
redshifts estimated by COMBO-17. We see that
the photometric redshift errors are small (i.e.,  $|z_{\rm spec}$-$z_{\rm phot}|$\,$\ltapp$\,0.1)
and they are not statistically larger in the case of the MIPS sources.
The latter can be explained as follows.
 Mid-infrared space-borne and ground-based
observations of local LIRGs/ULIRGs reveal that the dust responsible
for the bulk of the IR luminosity 
 of those objects
originates from very compact regions
\citep{Soifer00,Soifer01,Charmandaris02b}. 
The effect of extinction in these dusty systems is
therefore very localized and is usually  not apparent in the global spectral
energy distribution of their optical counterparts \citep{Sanders96}.
A similar situation
likely pertains at higher redshifts up to at least
$z$\,$\sim$\,1. At $z$\,$\sim$\,0.7 for instance, ISOCAM~15\mic and
MIPS~24\micpa-selected galaxies are indeed associated with luminous
optical sources
characterized by a wide range of optical colors and morphologies
(\citealt{Flores99,Rigopoulou02,Franceschini03}, Bell et al.  2005;
 see also Sect.\,6). Only a careful approach
based on medium-resolution spectroscopy can distinguish them from the
optically-selected galaxy population \citep[e.g.,][]{Flores04}. It is
therefore unlikely to encounter any significant increase of redshift
mis-classification {\it as a function of infrared luminosity\,} (at
least up to $z$\,$\sim$\,1).

\begin{figure*}[htpb]
\includegraphics*{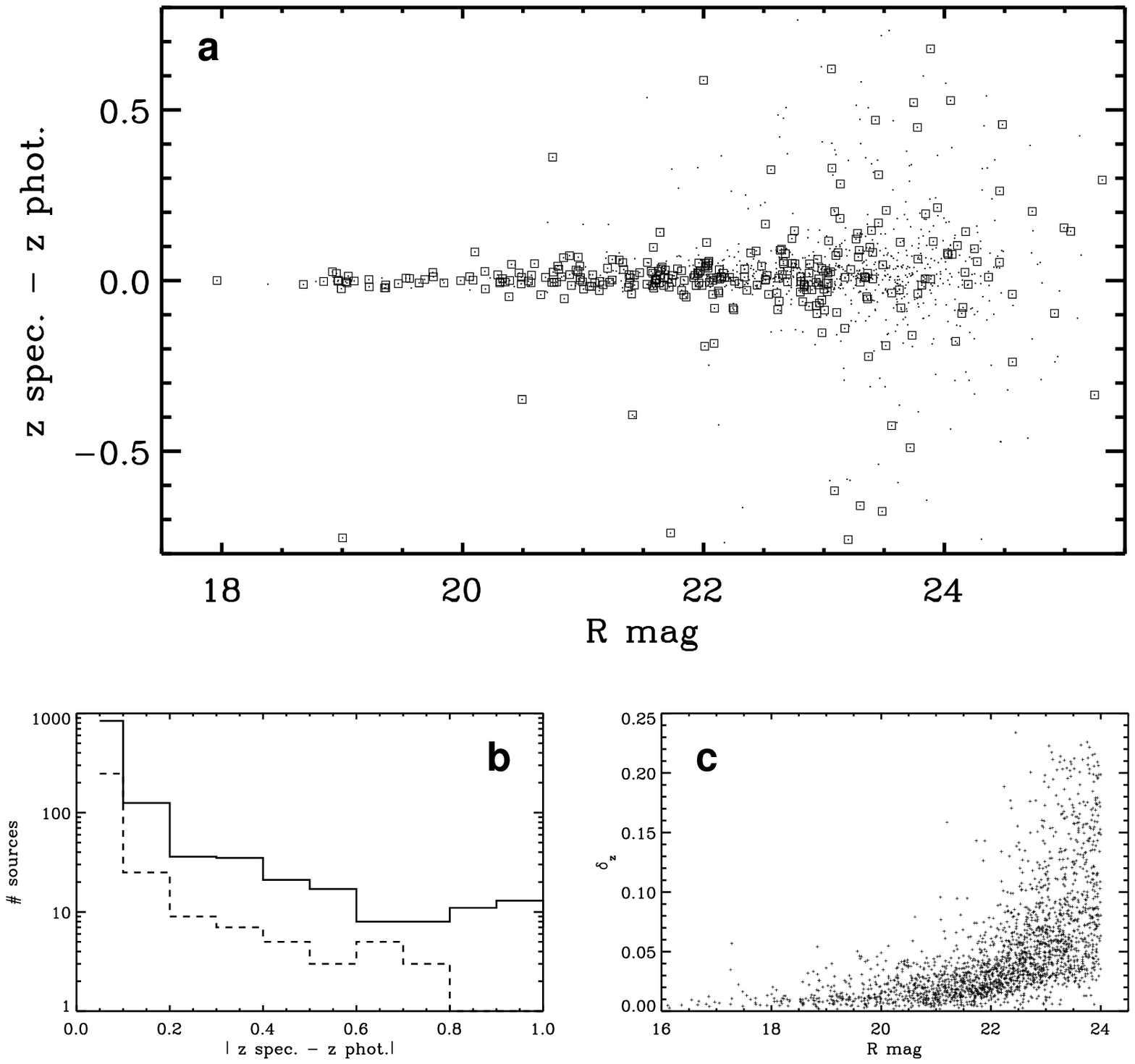}
\vskip 0.2cm \figcaption{ {\it a)\,} Comparison between the
VVDS spectroscopic  and COMBO-17 photometric redshifts as a function of
the $R$-band magnitude, for 1\,142 optically-selected galaxies (dots)
and 308 MIPS 24\mic sources  (open squares).
{\it b)\,} Histogram of the
absolute values $|z_{\rm spec}$-$z_{\rm phot}|$, with the optically-selected and 24\mic
sources respectively indicated by the solid and dashed lines.
{\it c)\,} Photometric redshift uncertainties provided by COMBO-17
 for the 24\mic sources  considered
in this paper.
Selection cut is set to
$R$\,$\leq$\,24\,mag and $z$\,$\leq$\,1.2. In agreement with comparisons
from panels  {\it a)\,} and  {\it b)}, the accuracy is
better than  $\delta_{z}$\,$\sim$\,0.1 for 88\% of the sample.}
\vskip .2cm
\end{figure*}

Finally, 
 Fig.\,2c shows the photometric redshift
uncertainties provided by the COMBO-17 catalog
as a function of the observed $R$-band magnitude for the MIPS
sources considered in this paper. These errors were determined 
as the square root of the variance characterizing the
redshift probability distribution of each object. In agreement
with the  comparison we made from Fig.\,2a using the
 spectroscopic sub-sample,
we see that sources brighter than $R$\,$\sim$\,22\,mag have a very
accurate redshift estimate (i.e., $\delta_{z}$\,$\ltapp$\,0.04)
and most of the sample (88\%) is characterized by a
redshift accuracy better than $\delta_{z}$\,$\sim$\,0.1. 
The
average uncertainty is only $<$$\delta_z$$>$=0.07 
with a dispersion   $\sigma_z$=0.05 for
sources with 22\,mag\,$\leq$\,$R\,$\,$\leq$\,24\,mag.
It rises to  $<$$\delta_z$$>$=0.09 (with a similar dispersion)
if we only consider objects fainter than $R$\,$\sim$\,23\,mag.
Looking at
the evolution of the MIPS sources averaged per redshift bin of
$\Delta$$z$\,$\sim$\,0.2
 should lead therefore to rather robust results.

\subsection{Completeness of the sample}

The biggest concern affecting
 our  results is
 the incompleteness of the redshift determination for the most distant
sources considered in this paper.
As previously mentioned, 3\,616 objects were detected by MIPS above
the 80\% completeness limit in the region overlapping with the optical
surveys of the CDFS, while our final catalog only contains 1\,962
sources with $f_{\rm 24\mu m}$\,$\geq$\,83\muJy and  a
redshift identification
(i.e., $\sim$\,55\%).  To better assess this effect, we
plot in Figure~3.a the fraction of MIPS sources
characterized either by a spectroscopic or a photometric redshift as a
function of the observed 24\mic flux (solid line) as well as the
fraction of their identification with an optical counterpart for
various magnitude limits (shaded regions).
At the highest fluxes (i.e., $f_{24\mu m} \gtapp$ 1mJy), the incompleteness of
the sample is mostly due to a few objects detected close to very
bright stars blooming the optical image and where a reliable
identification of the counterpart cannot be obtained.
These cases should not introduce any bias in our results.
Below 1\,mJy, the drop in the redshift determination completeness
corresponds to sources fainter than
$R$\,$\sim$\,23.5\,mag (see also Fig.\,3.b).
Above the 80\% completeness limit of our MIPS data (i.e., 83\muJypa),
we see  for instance that a significant fraction
($\sim$\,25\%) of the MIPS objects with no redshift are associated with 
24\,mag\,$\ltapp$\,$R$\,$\ltapp$\,25.5\,mag sources, while $\sim$\,20\% of them
have counterparts fainter than $R$\,$\gtapp$\,25.5\,mag.
 This also
indicates that a complete study of the faint MIPS population
will likely require an extensive use of photometric redshift
techniques. 

The consequence of this limitation can be seen by comparing
the redshift determination completeness as a function of the $R$-band magnitude with
the distribution of the $R$-band magnitudes of the MIPS sources
for different redshift limits. This is shown in Fig.\,3.b, while
Fig.\,3.c represents the absolute $R$-band magnitudes of the MIPS
sources as a function of redsfhit and derived from COMBO-17.
We see that the
identification should be nearly complete up to $z$\,$\sim$\,0.8, but
we start missing 24\mic objects located at higher redshift
and associated with
optical counterparts fainter than $R$\,$\sim$\,23.5--24\,mag.  
These mis-identifications  likely induce a biais against the faintest
sources detected at 24\micpa. In Sect.\,6 we will establish a
correlation between the optical and the infrared luminosities of galaxies in
our sample, and we will use such relation in Sect.\,7 to quantify how
this bias affects the estimates of luminosity functions (see also the 
Appendix for further details characterizing these faint 24\mic
sources). However,
given 
the importance of the cosmic
variance arising from the variations of large scale structures in the CDFS
 (see Sect.\,3.4), and taking into account
the errors in the $k$-corrections used to derive IR~luminosities (see Sect.\,5),  we 
 infer that the possible missing
redshift determinations
should not dominate the {\it absolute} uncertainty in the source density
estimate up to $z$\,$\sim$\,1.

\begin{minipage}[htbp]{8.5cm}
\vskip .3cm 
\centerline{\psfig{file=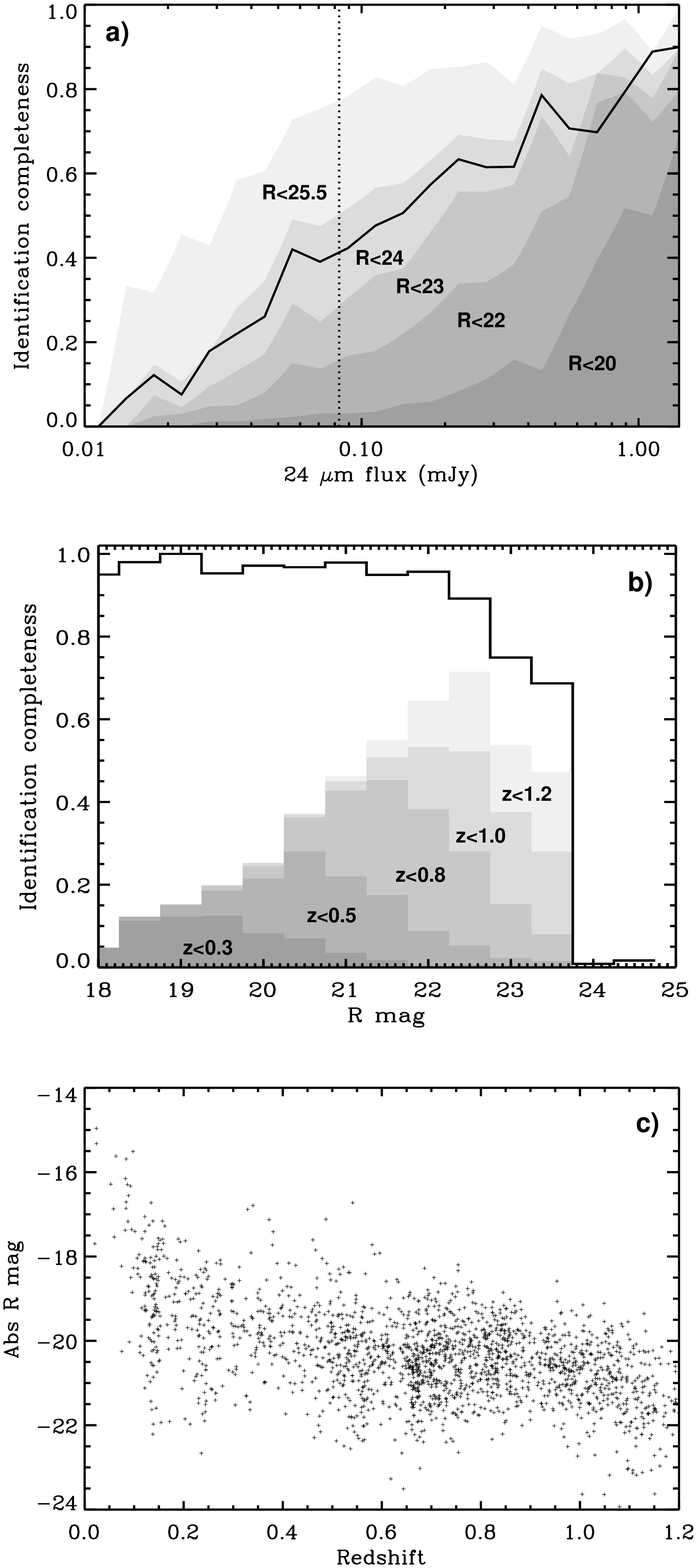,width=8.5cm,angle=0}}
\vskip 0.2cm \figcaption{{\it a)\,} Fraction of MIPS sources
identified with a spectroscopic or a photometric redshift as a
function of the observed flux at 24\mic (solid line).  We also
indicate the fraction of MIPS sources with an optical counterpart
detected in the $R$-band for various magnitude limits (shaded
regions).  The vertical dotted line corresponds to the 80\%
completeness limit of the 24\mic observations.  {\it b)\,} 
Fraction of 
 24\mic source optical counterparts
identified with a redshift,
as a function of the $R$-band magnitude (solid line). Shaded regions
show the $R$-band magnitude histograms (all scaled with an arbitrary
constant factor) of the MIPS sources up to various redshift
limits. The drop in the redshift identification is clearly apparent at
$R$\,$\sim$\,23.5-24\,mag. The sample should be complete up to
$z$\,$\sim$\,0.8.
{\it c)\,} Absolute $R$-band magnitudes of the MIPS sources as a function
of redshift.
}\vskip .2cm
\label{fig:z_distrib}
\end{minipage}

Other arguments
also suggest that the bulk of the MIPS sources
for which  we could not identify any reliable redshift should be indeed located
at $z$\,$\gtapp$\,1 and will not affect this study. First
we will show in the following sections that infrared luminous
galaxies are associated with  optical counterparts at the bright end
of the luminosity function in the visible.
Luminous optical sources with L\,$\gtapp$\,L$\star$
but fainter than $R$\,$\sim$\,23.5--24\,mag
should   lie at $z$\,$\gtapp$\,1 \citep{Benitez00}.
Moreover,
preliminary results from $Spitzer$  reveal that a
significant fraction of the MIPS  sources are located at such
large redshifts
 \citep{Chary04,Egami04a,LeFloch04,Lagache04}.
Finally,  although they are not 
 fully reliable, the COMBO-17 redshift estimates based on the peak
of the redshift probability distributions  indicate
that the majority of the sources that we have not identified
should be at $z$\,$\gtapp$\,1. 

\subsection{Redshift distribution}

Figure\,4 shows the distribution of the  redshifts derived for
our sample of infrared sources (bottom panel, solid line). It
is compared to the redshift distributions of the VVDS (dotted line)
and the COMBO-17 objects with $R \leq 24$ (dashed line). 
As previously discussed, the redshift identification of the
MIPS sources is  complete only up to
$z$\,$\sim$\,0.8, and the distribution beyond this limit should thus be considered
as a lower limit estimate.

\begin{minipage}[b]{8.5cm}
\vskip .3cm 
\centerline{\psfig{file=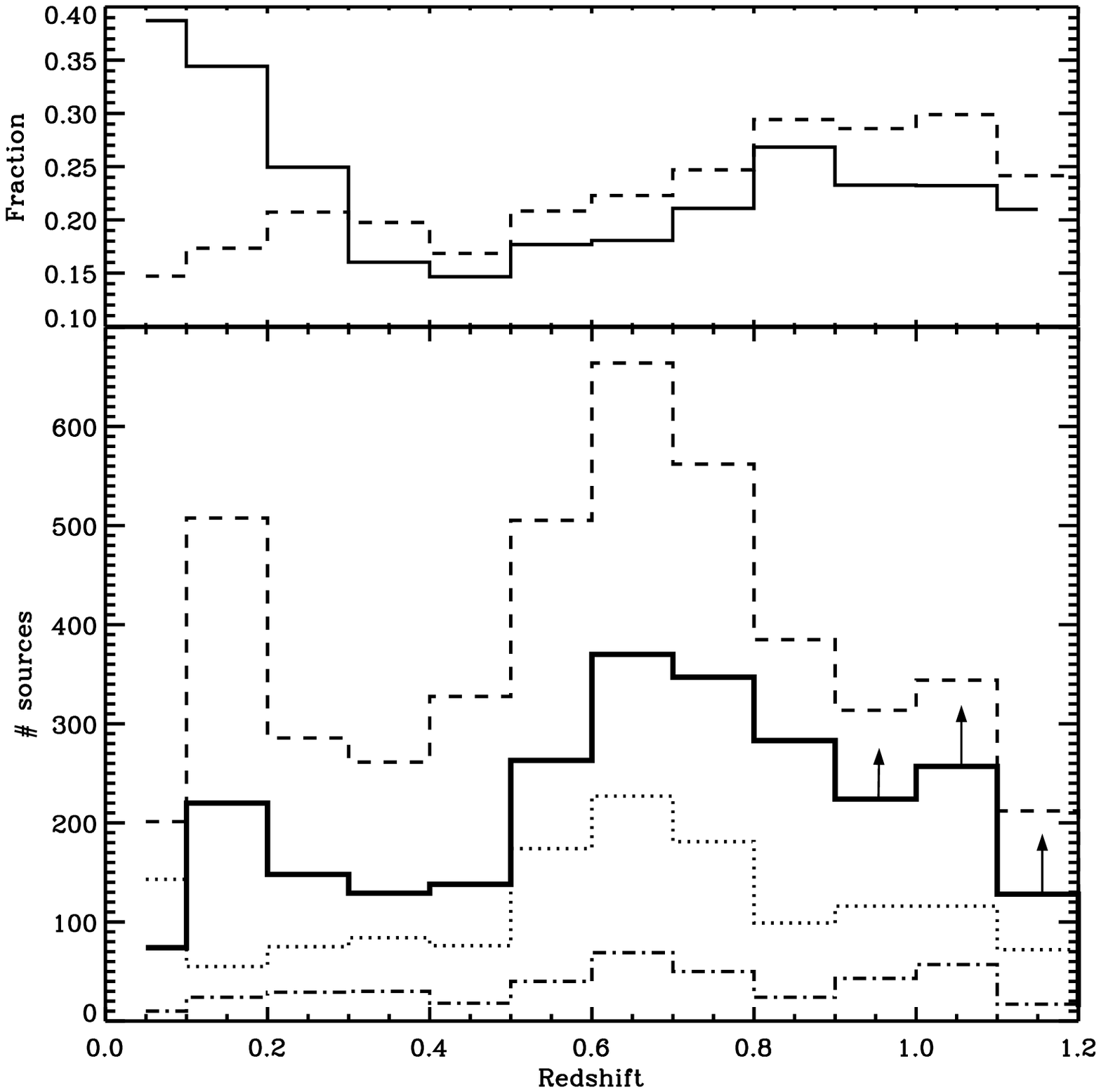,width=8.5cm,angle=0}}
\vskip 0.2cm \figcaption{{\it Bottom:\,} The redshift distribution of
  24\micpa-selected sources over a 775\,arcmin$^2$ region within the
  CDFS (solid line), compared to the distributions of photometric
  redshifts for $R$\,$\leq$\,24\,mag sources in the COMBO-17 catalog
  (dashed-line, scaled down by a factor of 2.5) and spectroscopic
  redshifts for $I_{AB}$\,$\leq$\,24\,mag sources in the VVDS survey
  (dotted line).  The contribution of MIPS sources with spectroscopic
  redshifts is also indicated (dash-dotted line). Note the prominent
  overdensity at $z$\,$\sim$\,0.65, clearly seen in all the
  distributions. Source densities at $z$\,$\geq$\,0.8 should be
    considered as lower limit estimates due to the incompleteness of
    the redshift identification.  {\it Top:\,} The
  fractions of $R$\,$\leq$\,24\,mag sources detected at 24\micpa,
  estimated as a function of redshift in the whole population of
  COMBO-17 (dashed line) and restricted to the objects with M$_{\rm
    B}$\,$\leq$\,--16 (solid line).
}\vskip .2cm
\label{fig:z_distrib}
\end{minipage}

We note that the three distributions (i.e., MIPS, VVDS, COMBO-17) are
more or less similar from $z$\,$\sim$\,0 to $z$\,$\sim$\,1. 
Up to this redshift limit, 
MIPS is indeed
sensitive to sources luminous in the infrared but also to more normal
galaxies (see Sect.\,5) such as those
 detected at optical wavelengths and driving
the VVDS and COMBO-17 redshift distributions.  
Furthermore, we note
a significant number of objects around
$z$\,$\sim$\,0.65, apparent in all the distributions.
This originates from an overdensity characteristic of
the CDFS near this  redshift \citep[e.g.,][]{Wolf04}
and it is likely   related to cosmic variance and large-scale structure effects 
\citep{Somerville04}.
It may constitute
a non negligible source of uncertainty in our density
estimates.
A comparison between the integrated $B$-band luminosity densities as a
function of redshift from galaxies of the blue sequence
(Bell, private communication, see Willmer et al., in prep., for the
evolution of the blue sequence luminosity function)
in the CDFS and averaged over the three fields
of COMBO-17 \citep{Wolf03} shows that the CDFS is most often
underdense by $\sim$50\% at 0\,$\ltapp z \ltapp$\,1, except indeed at
$z$\,$\sim$\,0.65$\pm$0.1 where the overdensity reaches 
$\sim$20\%. Given the similarity between the redshift distributions
of the optically and infrared selected sources, and
since most of MIPS sources at $z$\,$\sim$\,0.7 appear to be
associated with large spirals dominating the $B$-band emission (Bell
et al. 2005, see also Fig.\,10b),
this cosmic variance characterized from  the 
$B$-band luminosity density is likely affecting also the population
of galaxies detected at 24\micpa. Therefore the apparent peak at $z$\,$\sim$\,0.65
in the MIPS redshift distribution 
is probably only related to the structure of the CDFS and it should not be
 a characteristic of the general population of infrared-selected
sources.

In spite of the similarities that we observe 
between the redshift distributions
in the visible and the infrared, a more critical
look reveals that the fraction of optical sources
brighter than M$_{\rm B}$\,=\,--16
and detected at 24\mic decreases from $z$\,=\,0 to $z$\,$\sim$\,0.4 and then
increases with redshift up to $z$\,$\sim$\,1 
(see Fig.4., top panel).  Interpreting this trend is not
straightforward, 
but it could reflect the difference in the $k$-correction
effects observed at optical and infrared wavelengths. As we will see
in Sect.\,5 (see Figs.\,7 \&~9) the effective sensitivity of the deep
24\mic observations drops rapidly from $z$\,=\,0 to $z$\,$\sim$\,0.5, 
which explains the sharp decrease of relative 24\mic detections
in this redshift range. Then
 it remains  roughly constant at
0.5\,$\ltapp$\,$z$\,$\ltapp$\,0.9 because of strong emission features
present at 11.3 and 12.7\mic in galaxy spectra progressively entering
the 24\mic filter at these redshifts. Since the sensitivity of the optical
data constantly drops with distance,  an increase in the relative
number of 24\mic sources detected beyond $z$\,$\sim$\,0.5 in the visible
 might thus result. Note that it may also reflect the stronger evolution
of IR sources with lookback time, since IR-luminous phases in galaxies
were more common in the past \citep[e.g.,][]{Hammer05}.
 The lack of apparent decline moreover
suggests that the intrinsic peak of the MIPS population redshift
distribution must  lie at least beyond $z$\,$\sim$\,0.8.

Finally, another overdensity  lies at $z$\,$\sim$\,0.15. 
It represents a rather small fraction of our
detections at 24\mic (i.e., 4\%). Since we will only consider
sources at $z$\,$\geq$\,0.3 when building the luminosity functions
in Sect.\,7, it will not affect our final results on the evolution
of the IR~galaxy population.

\section{Exploring the origin of the break in the MIPS 24\mic number counts}

\subsection{Cumulative differential  counts as 
a function of redshift}

Using our matched catalog of 24\micpa-selected objects with
redshifts, we now explore for various redshift slices
the contribution of these
MIPS sources to the total differential number counts.
Our goal is a better understanding of the
origin of the bump detected at 0.2--0.4\,mJy
\citep[P04,][]{Marleau04}. 
Figure\,5 illustrates these cumulative source counts for 
redshift limits of 0.3, 0.5, 0.8, 1.0 and 1.2
along with
 the global contribution of the MIPS sources identified with an optical
counterpart in the full COMBO-17 catalog (i.e., $R$\,$\ltapp$\,25.5\,mag).
The distributions for redshift limits $z$\,=\,1 and $z$\,=\,1.2 are
likely underestimated due to the incompleteness of
the sample as showed in Sect.\,3.3.  However,
uncertainties due to cosmic variance likely dominate here
($\sim$\,0.10-0.15\,dex based on the ratio between the $B$-band luminosity
densities of the CDFS and averaged within the other fields of
COMBO-17).
For comparison, we also represent the total 24\mic
number counts derived from our sub-sample as well as
those obtained by \citet{Papovich04}, who averaged the 24\mic source
population over $\sim$\,10\,deg$^2$ in several cosmological fields. 
For consistency with our data, these counts from P04
 were not corrected for  incompleteness of the MIPS detections at
faint fluxes.

\begin{minipage}[htb]{8.5cm}
\vskip .3cm 
\centerline{\psfig{file=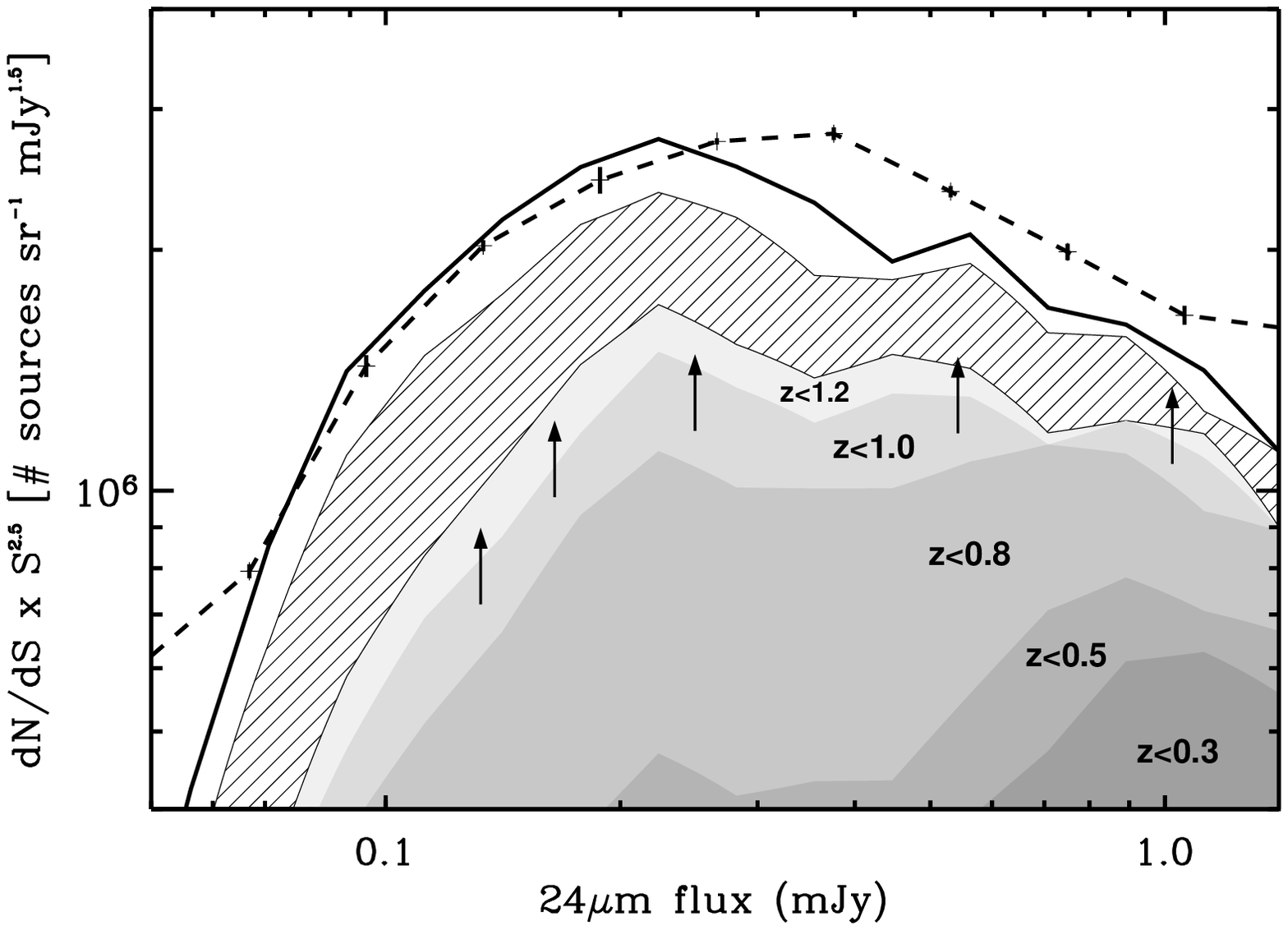,width=8.5cm,angle=0}}
\vskip 0.2cm \figcaption{Cumulative 
differential number counts observed at 24\mic as a function of redshift
(shaded regions).
Distributions at $z$\,$\geq$\,0.8
should be considered as lower limit estimates due to the incompleteness of the
redshift identification.
The striped area
represents the contribution of all 24\mic sources identified with
an optical counterpart down to $R$\,$\ltapp$25.5\,mag.  The total
differential
counts derived from our sub-sample are illustrated by the black solid
line, while those determined by \citet{Papovich04}
are represented by the
dashed line and the vertical error bars.
Counts are normalized to the
Euclidian slope.}
\vskip .2cm
\label{fig:z_distrib}
\end{minipage}

We note that the 24\mic number counts derived  in the CDFS
show a roughly good agreement
with those obtained by P04 below $f_{24\mu m}$\,$\sim$\,0.2\,mJy.
However, the
density of 24\mic CDFS sources at higher fluxes
appears 
to be slightly lower ($\sim$20\%),
and consequently the peak of the differential number counts
seems to occur at  fainter levels.  As previously seen in
Sect.\,3.4., such variations are easily understood in terms of cosmic
variance and the smaller number of infrared-bright sources in the CDFS
 is likely related  to the underdensity observed
in the $B$-band at  $z$\,$\ltapp$\,1. 

At critical fluxes $f_{24\mu m}$\,$\sim$\,0.2-0.4\,mJy
 where the 24\mic differential
counts present a turn-over, sources lying at $z$\,$\ltapp$\,1
contribute $\sim$\,55\% to the whole sample.  Even though this
fraction could be slightly higher due to the possible incompleteness of 
our redshift identification at 0.8\,$\ltapp$\,$z$\,$\ltapp$\,1,
a significant part of
the MIPS population should  therefore be located at higher redshifts
(i.e., $z$\,$\gtapp$\,1). 
As already derived from Fig.\,3, it can also
be noted that
$\sim$\,20\% of the 24\mic sources have
optical counterparts fainter than $R$\,$\sim$\,25.5\,mag.

\subsection{Comparison with model predictions}

In Figure\,6 we compare the  differential counts in four redshift slices
between $z$\,=\,0 and $z$\,=\,1.2 
 with the predictions from the  models of IR
galaxy evolution proposed by \citet{Lagache04}, \citet{Chary04},
 \citet[, see also Pozzi et al. 2004]{Gruppioni05}
and \citet{Pearson05}.
These models are all tied
to the local IR galaxy population but differ
in (i) the description of its components and their 
global properties (i.e., SEDs, luminosity
functions per object type)
 and (ii) the assumptions and parameterization
used to infer the backward evolution of IR sources.
They provide a reasonably good
fit to the total  24\mic number counts. They also reproduce
 a variety of other observables such as the IR background,
the counts and/or the redshift distributions of the $ISO$ 
and SCUBA galaxy populations. 

\begin{minipage}[htb]{8.5cm}
\vskip .3cm 
\centerline{\psfig{file=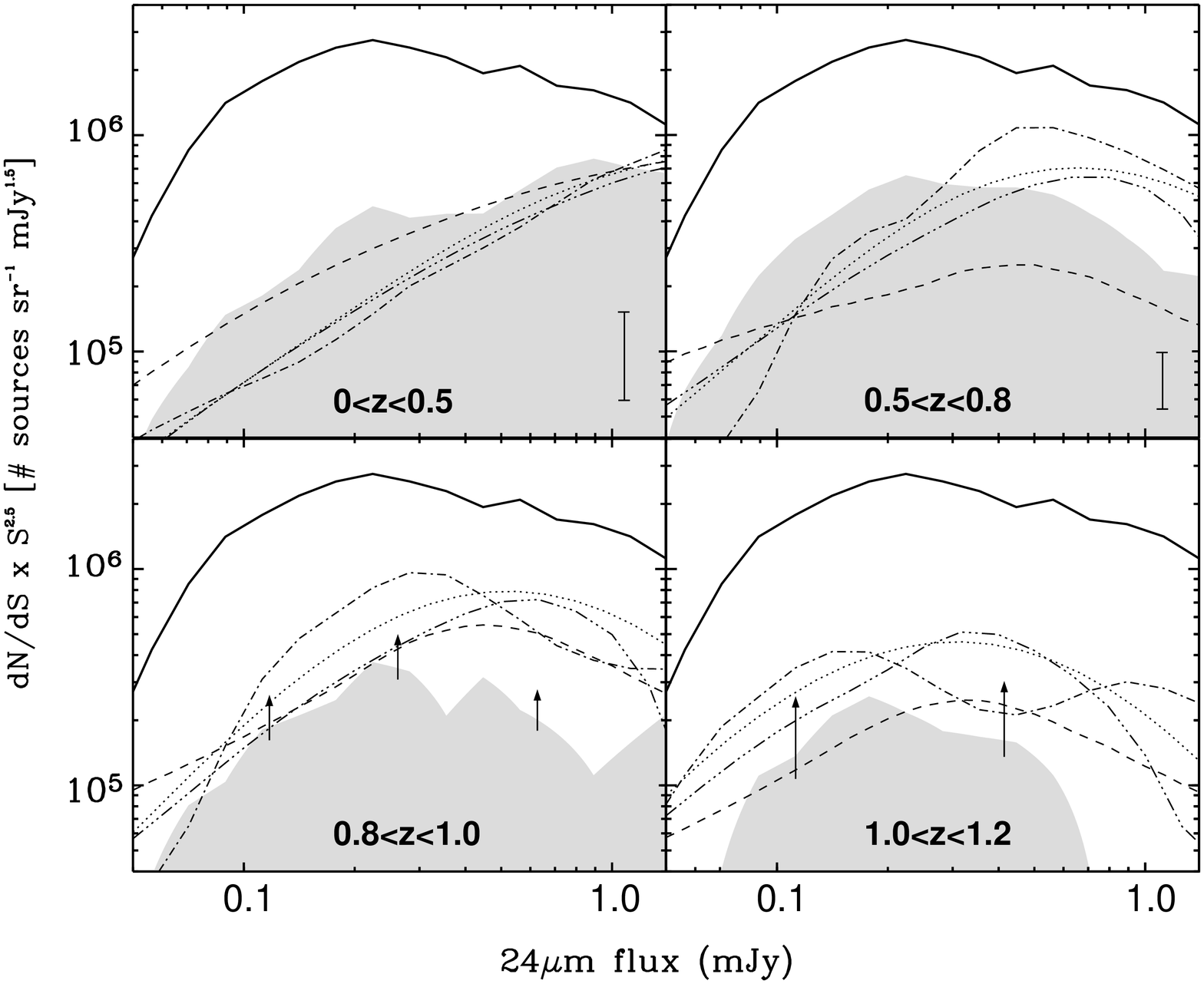,width=8.5cm,angle=0}}
\vskip 0.2cm \figcaption{ 
Differential 24\mic number counts produced by the MIPS sources
in four redshift slices between $z$\,=\,0 and $z$\,=\,1.2
as indicated within each panel (shaded regions). They are
 compared to the predictions by \citet[, dashed lines]{Lagache04},
\citet[, dotted lines]{Gruppioni05}, \citet[, dash-dotted lines]{Chary04}
and \citet[, triple dot-dashed line]{Pearson05} 
for
similar redshift ranges. The total counts derived from our sample are shown
in each panel by the black solid line. The vertical bar in the upper panels
represents the typical uncertainty due to the effect of large scale structures.
The counts at 0.8\,$\leq$\,$z$\,$\leq$\,1.0 and 1.0\,$\leq$\,$z$\,$\leq$\,1.2
(lower panels) only show a lower limit  given the incompleteness
of the redshift identification in our sample at $z$\,$\gtapp$\,0.8.
}
\vskip .2cm
\label{fig:z_distrib}
\end{minipage}

Looking at their predictions  for the 24\mic number
counts within the
 redshift slices considered  in this Fig.\,6, we see that
the models from \citet{Chary04}, \citet{Gruppioni05} and \citet{Pearson05}
may underestimate the
 contribution of faint sources at 0\,$\ltapp$\,$z$\,$\ltapp$\,0.8 
but predict too many bright objects at $z$\,$\gtapp$\,0.5.
\citet{Lagache04}, on the other hand, closely follow the observations
 at $z$\,$\ltapp$\,0.5 but might globally underestimate
the source density at 0.5\,$\ltapp$\,$z$\,$\ltapp$\,0.8.
All these models finally seem to overpredict the contribution of bright galaxies
at  $z$\,$\gtapp$\,0.8. However we also note  that 
 these comparisons could be undermined given not only cosmic variance effects and
the global
 underdensity of the CDFS between $z$\,=\,0 and $z$\,$\sim$\,1 but also the
 incompleteness of our redshift identification at $z$\,$\gtapp$\,0.8.

Comparisons with other models
in the literature such as those published before the launch
of $Spitzer$ lead to  larger discrepancies. These
 models tend to predict the characteristic turn-over of
the differential 24\mic number counts at much higher fluxes
than where it has been observed (see for instance Figure\,3 of
\citealt{Papovich04}).

\section{Total infrared luminosities}

\subsection{Estimating the total IR luminosity of galaxies from their mid-IR emission}

The rest-frame mid-infrared (MIR) regime (i.e., 8\mic$\ltapp \lambda
\ltapp$40\micpa) is considered to be a  good tracer of the
bolometric luminosity of galaxies. Using the 12\mic IRAS galaxy
sample, \citet{Spinoglio95} first showed that the relative dust
content of galaxies balances their total energetic output between the
optical and the far-infrared, leaving a ``pivot point'' in the MIR
where the specific luminosity linearly scales with the bolometric
one. More recently, a similar conclusion has been reached by
\citet{Chary01} who found a tight correlation between the 15\mic and
the total infrared luminosities for a sample of normal and luminous
sources studied with ISOCAM. Based on
these considerations and taking into account  the general IR/submm color-color
and/or luminosity-color correlations  observed  in the
local Universe,
several
authors have built libraries of luminosity- or color-dependent
galaxy  templates,
which can then be used to estimate the total infrared luminosity of
galaxies from their 24\mic flux densities
\citep{Dale01,Chary01,Dale02,Lagache03,Chanial03}.

Figure\,7 shows the total infrared luminosity that can be detected in
our survey down to a 24\mic sensitivity limit of 0.08\,mJy, as a
function of redshift and computed with the aforementioned
libraries. Provided these templates are still representative of high
redshift sources, we see that MIPS can easily detect normal starbursts
(i.e., L$_{\rm IR}$\,$\ltapp$\,10$^{11}$\,L$_{\odot}$) up to
$z$\,$\sim$\,1, and LIRGs up to $z$\,$\sim$\,1.5. Note that such
limits only characterize a pure selection of sources at 24\micpa, and
the use of {\it a priori\,}
information from detections at other wavelengths can allow the
identification  of MIPS sources with similar
infrared luminosities  at even higher
redshift \citep[see e.g.,][]{LeFloch04}. The 80\% completeness level 
 roughly corresponds to a 5$\sigma$ detection in the MIPS
data \citep{Dole04a}, and the quoted 0.08\,mJy limit
is therefore rather conservative.

Figure\,7 also shows the dispersion 
between the predictions (and therefore the
templates) of the various  libraries \citep[see also][]{Papovich02}.
For a given observable (e.g., infrared luminosity, color,~...)
these libraries assign  a unique SED
that slightly varies 
from one set of templates to another.
Such variations  reflect an intrinsic dispersion of the different correlations
observed between the MIR/FIR/submm properties of galaxies that is
usually not taken into account \citep[but see][ for the use of
bi-variate luminosity functions addressing this issue]{Chapman03a}.
As an example, we compare in Figure\,8 the observed relation between L$_{\rm
IR}$ and the monochromatic 15\mic luminosity L$_{15}$\footnote{We
define the monochromatic luminosity as L$_\nu$\,=\,$\nu \times S_\nu$, 
with $S_\nu$  the monochromatic flux of the galaxy expressed in W\,Hz$^{-1}$.}
 for a sample of local galaxies studied with $ISO$ and IRAS
\citep{Chary01}
with the predictions of the template libraries considered in Fig.\,7.
We see that the dispersion can easily reach $\sim$\,0.2\,dex.

\begin{minipage}[htb]{8.5cm}
\vskip .3cm 
\centerline{\psfig{file=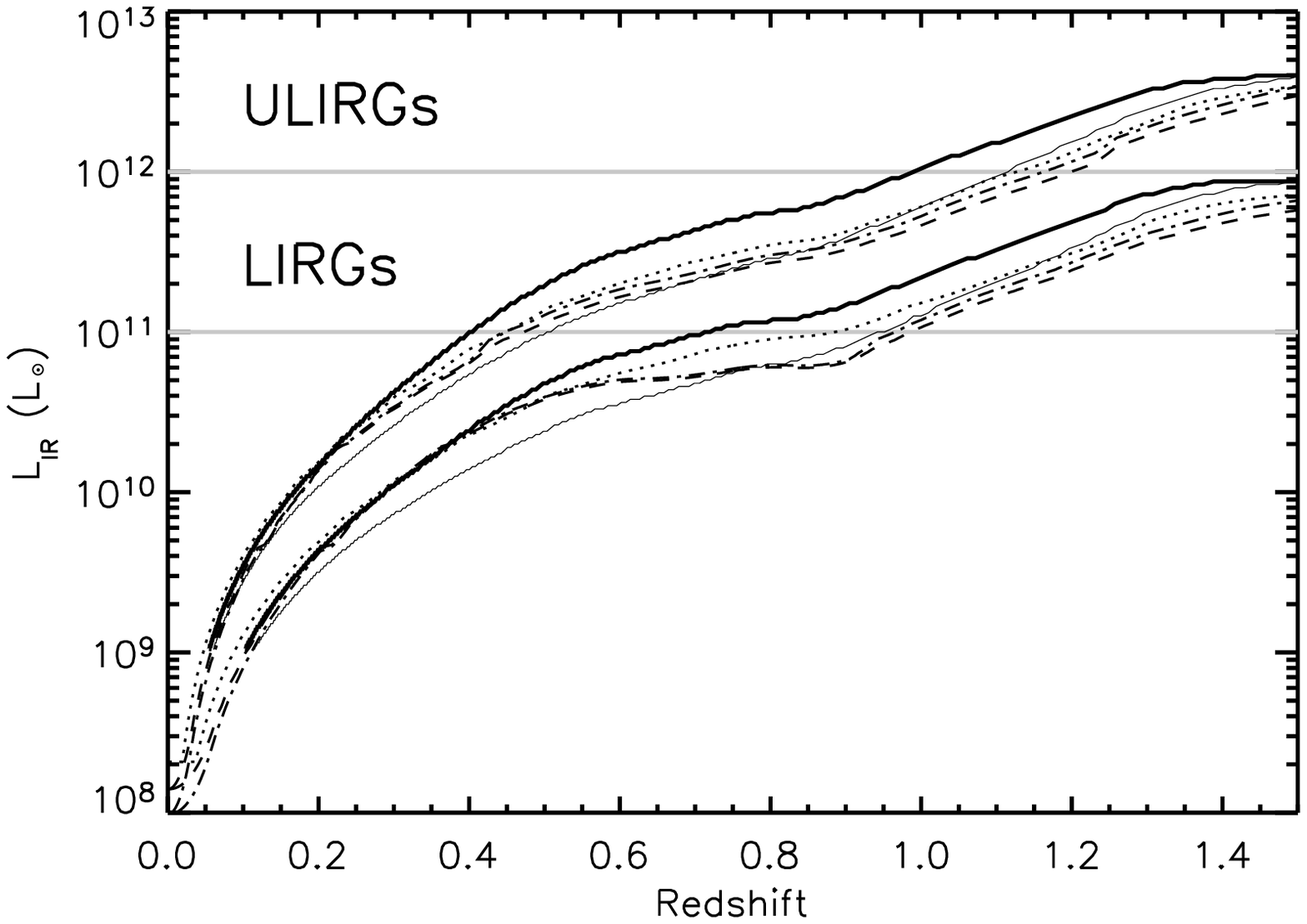,width=8.5cm,angle=0}}
\vskip 0.2cm \figcaption{
 Detection limits in terms of total infrared luminosity, derived for 
two different MIPS 24\mic sensitivities and assuming
the template libraries of \citet[, thick solid line]{Lagache04}, 
\citet[, thin solid line]{Lagache03}, \citet[, dotted line]{Dale01}, 
\citet[, dashed line]{Chary01} and \citet[, dash-dotted line]{Chanial03}
as a function of redshift.
The lower (respectively, upper) set of curves corresponds to a
flux limit of 0.08\,mJy (0.3\,mJy) typical of the MIPS deep
(shallow) surveys. 
}\vskip .2cm
\label{fig:z_distrib}
\end{minipage}

To explain this dispersion, one may note that in
 the local Universe the
mid-infrared SEDs of starburst sources with comparable bolometric
luminosities are subject to significant variations 
\citep[e.g.,][]{Armus04}. In the rest-frame wavelength
range probed by the MIPS 24\mic filter for high redshift galaxies,
the MIR emission results  from the combination of prominent
broad-band features mostly observed between 3 and 14\mic and usually
denoted the Polycyclic Aromatic Hydrocarbons bands (PAHs),
superimposed on a rising continuum of Very Small Grains (VSGs)
stochastically heated by the young star radiation field \citep[see
e.g.,][ and references therein]{Laurent00}.  PAHs present an
amazingly universal SED signature as a global set of features when
integrated over normal spiral galaxies (\citealt{Roussel01}, but see
\citealt{Smith04} for smaller scale variations and a newly-discovered
17.1\mic PAH in NGC\,7331).
However, they are not detected in low metallicity sources
\citep[; Engelbracht et al. 2005, submitted]{Thuan99,Houck04b}. They are also believed to be destroyed
within intense radiation fields as in the vicinity of active galactic
nuclei \citep[AGN, e.g.,][]{LeFloch01}. Finally, the temperature of
the VSGs as well as the silicate absorptions at 9.7\mic and 18\mic
represent other important factors shaping the underlying continuum
of the global mid-infrared SED of galaxies. All of these variations
can  therefore bring significant uncertainties in relating the MIR
emission to the full infrared range of galaxy SEDs.

\begin{minipage}[htb]{8.5cm}
\vskip .3cm 
\centerline{\psfig{file=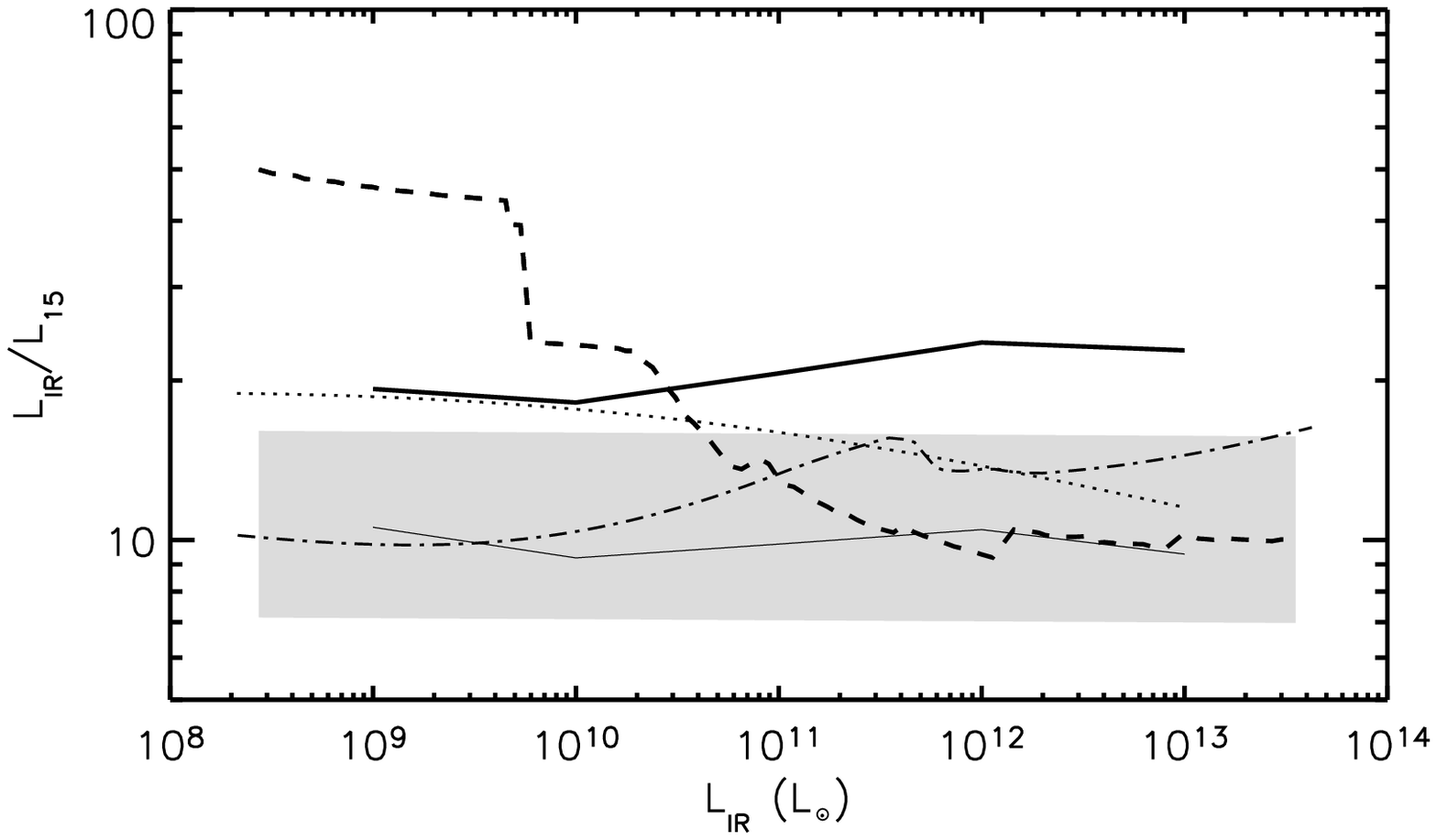,width=8.5cm,angle=0}}
\vskip 0.2cm \figcaption{
 Ratio between the total IR and the monochromatic
15\mic luminosities as a function of L$_{\rm IR}$ predicted by the SED libraries
considered in Fig.\,7
 (similar style coding). The grey area represents
the relation (mean value with 1$\sigma$
boundaries) derived by \citet{Chary01} for galaxies that were observed with
 ISO and IRAS.
}\vskip .2cm
\label{fig:z_distrib}
\end{minipage}

Furthermore, one may question whether these templates derived from the
properties of local galaxies are trully representative of higher
redshift sources.  Infrared luminous galaxies detected by  ISOCAM
 at a median redshift of $\sim$\,0.7 are  characterized by
roughly half-solar metallicites \citep{Liang04}, which could point to
IR SEDs slightly different from those of local LIRGs. However, PAHs
are still detected at this metallicity range in nearby starbursts
(Engelbracht et al. 2005, submitted). Moreover, the 24/15\mic
flux ratios observed in $z$\,$\sim$\,0.7 ISOCAM sources still argue
for the presence of significant PAHs in their SEDs \citep{Elbaz05}, 
and prominent PAH features were also recently
observed with the InfraRed Spectrograph  on {\it
Spitzer\,} \citep[IRS,][]{Houck04a} at even higher redshifts \citep[i.e.,
1.7\,$\leq$\,$z$\,$\leq$\,2.8,][]{Houck05}. 
We also note  that the MIR/radio relation expected from
the MIR/FIR and FIR/radio correlations observed locally
\citep{Condon92,Chary01} appears to be still valid in the more distant
Universe \citep[e.g.,][]{Elbaz02b,Appleton04}.
These   results give reasonable
confidence in using  local galaxy templates to estimate the 
infrared luminosities of distant IR sources.

\subsection{The total infrared luminosity of the MIPS sources at $z$\,$\ltapp$\,1}

For each library of SEDs previously discussed, we computed the redshift-dependent
relations between L$_{\rm IR}$ and the flux density observed at 24\micpa.
For every  source of the sample, 
an estimate of the total IR luminosity 
 was
thus obtained from each set of templates.
 These measures
were weigthed based on the accuracy of their associated library to
reproduce the
observed relation between  L$_{\rm
IR}$ and L$_{15}$  at the derived luminosity (see Fig.\,8). They were combined
 to provide the final infrared luminosity of the object, and their
associated rms was taken as an estimate of the corresponding
uncertainty.  

As noted by \citet{Chapman03a} there might also be a bias of
 MIPS 24\mic sources toward  luminous galaxies with hot dust
temperatures compared to the whole population of IR-luminous objects
at high redshift. This could result in additional systematics
affecting these luminosity estimates, and we have not formally
considered this effect here. We believe however that the uncertainties
due to  the
 SED libraries dominate the
 systematics in our study.

 The IR luminosities are displayed as a function of redshift in
Figure~9. Based on the previous arguments we estimate that they are
accurate within a factor of $\sim$\,2--3 up to $z$\,$\sim$\,1. We see
that most sources detected below $z$\,$\sim$\,0.5 are only modest
infrared emitters (i.e., L$_{\rm
IR}$\,$\leq$\,10$^{11}$\,L$_{\odot}$), and the number of
infrared-luminous galaxies found  at these low redshifts is quite low. 
At
larger distances however, LIRGs represent a significant fraction
of
the MIPS galaxy sample in terms of detection number.
This large population of LIRGs at  0.5\,$\ltapp$\,$z$\,$\ltapp$\,1.0
has also been observed in  various surveys performed at 15\mic with 
ISOCAM \citep[e.g.,][]{Aussel99,Flores99,Elbaz02b}. 
We finally note that the most luminous galaxies (i.e., L$_{\rm
IR}$\,$\geq$\,10$^{12}$\,L$_{\odot}$) are still pretty rare,
but there is a clear hint for an
increase of the number  of the brightest  LIRGs 
 (5$\times$10$^{11}$\,L$_{\odot}$\,$\ltapp$\,L$_{\rm
IR}$\,$\leq$\,10$^{12}$\,L$_{\odot}$) at $z$\,$\sim$\,1, which could point
to a large population of ULIRGs at even larger distances (i.e.,
$z$\,$\gtapp$\,1.5, \citealt{Blain99b,Chapman03}).  These detection rates are obviously
contingent on the comoving volume sampled at each redshift though.
This general issue will thus be addressed in more detail
in Sect.\,7, where we characterize the evolution of the infrared
luminosity function based on this sample of MIPS
sources.

\begin{figure*}[htpb]
\includegraphics*{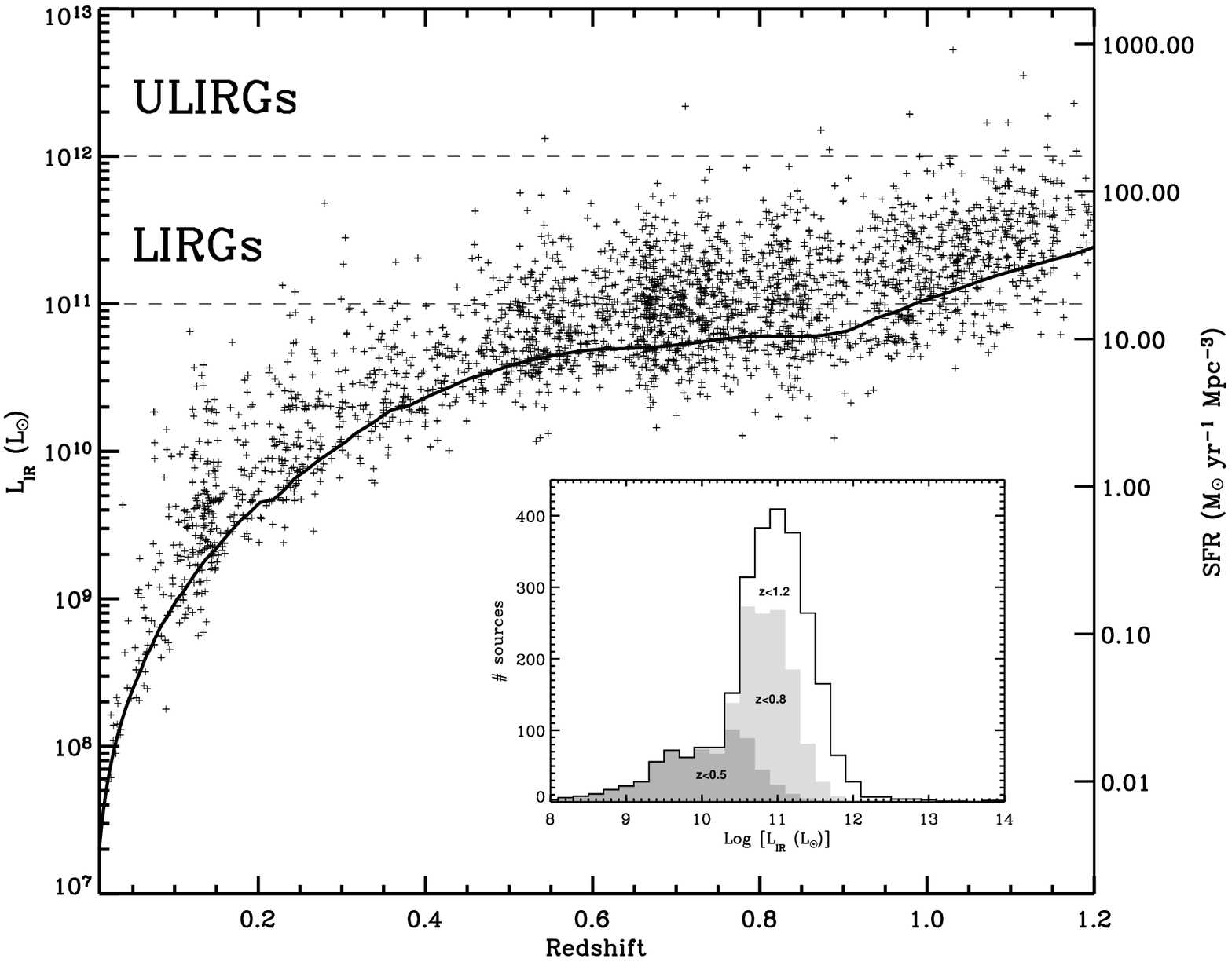}
\vskip 0.2cm \figcaption{
Total infrared luminosities of the MIPS 24\mic sources
identified with a redshift at 0\,$\leq$\,$z$\,$\leq$\,1.2 (+ symbols). They were
 derived using various
luminosity-dependent SED templates of the literature (see text
for more details) and we estimate that they
 are  accurate within a factor
of $\sim$\,2--3. The median uncertainty of the photometric
redshifts is less than 0.05 \citep{Wolf04}. Error bars are not shown for clarity.
Assuming the calibration from \citet{Kennicutt98},
IR luminosities are translated   into an 
``IR-equivalent SFR'' reported
 on the right vertical axis.
The thick solid line indicates as a function of redshift
the  infrared
luminosity corresponding to an observed 24\mic flux  
of 0.08\,mJy
(80\% completeness limit of our survey).
{\it Inset:\,} Corresponding IR luminosity histogram of the
sample, with the contribution of sources at $z$\,$\leq$\,0.5
(respectively,   $z$\,$\leq$\,0.8)    indicated
by the dark (light) shaded region.  }
\vskip .2cm
\end{figure*}

As a final remark, it is worth mentioning another possible caveat related to these 
infrared luminosity estimates.  The various template libraries that we
used in this work are only representative of normal and starburst-like
galaxies, and do not include the SEDs typical of active galactic
nuclei which are significantly flatter in the MIR wavelength range.  A
careful investigation of the nature of the emission process dominating
the 24\mic flux in our sub-sample (star-forming activity versus AGN)
is beyond the scope of this paper, but we note that
the fraction of MIPS sources showing evidence
for the presence of an AGN in their optical counterparts is less than $\sim$\,15\%
according to the VVDS and COMBO-17 
classifications. Furthermore,
recent synthetic models connecting the X-ray and
infrared SED of AGNs as well as their
contribution to the cosmic backgrounds in these energy bands indicate
that the emission arising from pure AGNs should be
negligible (i.e., $\ltapp$\,10\%) in high redshift sources 
detected at 24\mic \citep[e.g.,][]{Silva04}. In the local Universe moreover, 
AGNs dominate the mid-IR output of galaxies only for the most extreme ULIRGs
\citep[e.g.,][]{Lutz98,Tran01}, while 
the X-ray
and  optical spectral properties of infrared galaxies detected
with $ISO$ and MIPS at 15\mic and 24\mic 
also argue for a population dominated by star formation at the $\sim$\,85--90\% level
\citep{Fadda02,Franceschini03,Manners04,Franceschini05,Bell05}.  It should
therefore be reasonable to assume SEDs typical of star-forming
galaxies for this current work.

\section{The optical counterparts of the MIPS sources:
some properties}

We describe in this section a few basic characteristics of the
MIPS source optical counterparts  to provide a first-order
answer to very simple questions: are the distant infrared luminous galaxies
detected by MIPS
also luminous in the optical or are they
 highly obscured~? Are these dusty starbursts
associated with very young systems or already evolved galaxies~?
A  description of their morphologies, colors
 and specific star formation
rates is provided by Bell et al. (2005).

The COMBO-17 $R$-band magnitudes of the MIPS sources
are plotted as a function of the flux density at  24\mic in Figure\,10a. 
 Not surprisingly, there is a clear trend for the
fainter 24\mic objects to be associated with 
faint optical sources
likely located at higher redshifts. Such a  trend has already
been noted among the population of infrared
galaxies detected with ISOCAM at 15\mic \citep[e.g.,][]{Pozzi04}.
However there is  large dispersion in the relation 
 and for most of the sample
the
optical brightnesses can vary by more than 5 magnitudes
when infrared fluxes only change by less than a factor of 10.
This clearly points to a very wide range
 of the L$_{\rm IR}$/L$_{\rm optical}$
ratios and a broad variety in the nature
of the MIPS sources (see also the Appendix).

\begin{figure*}[htpb]
\includegraphics*{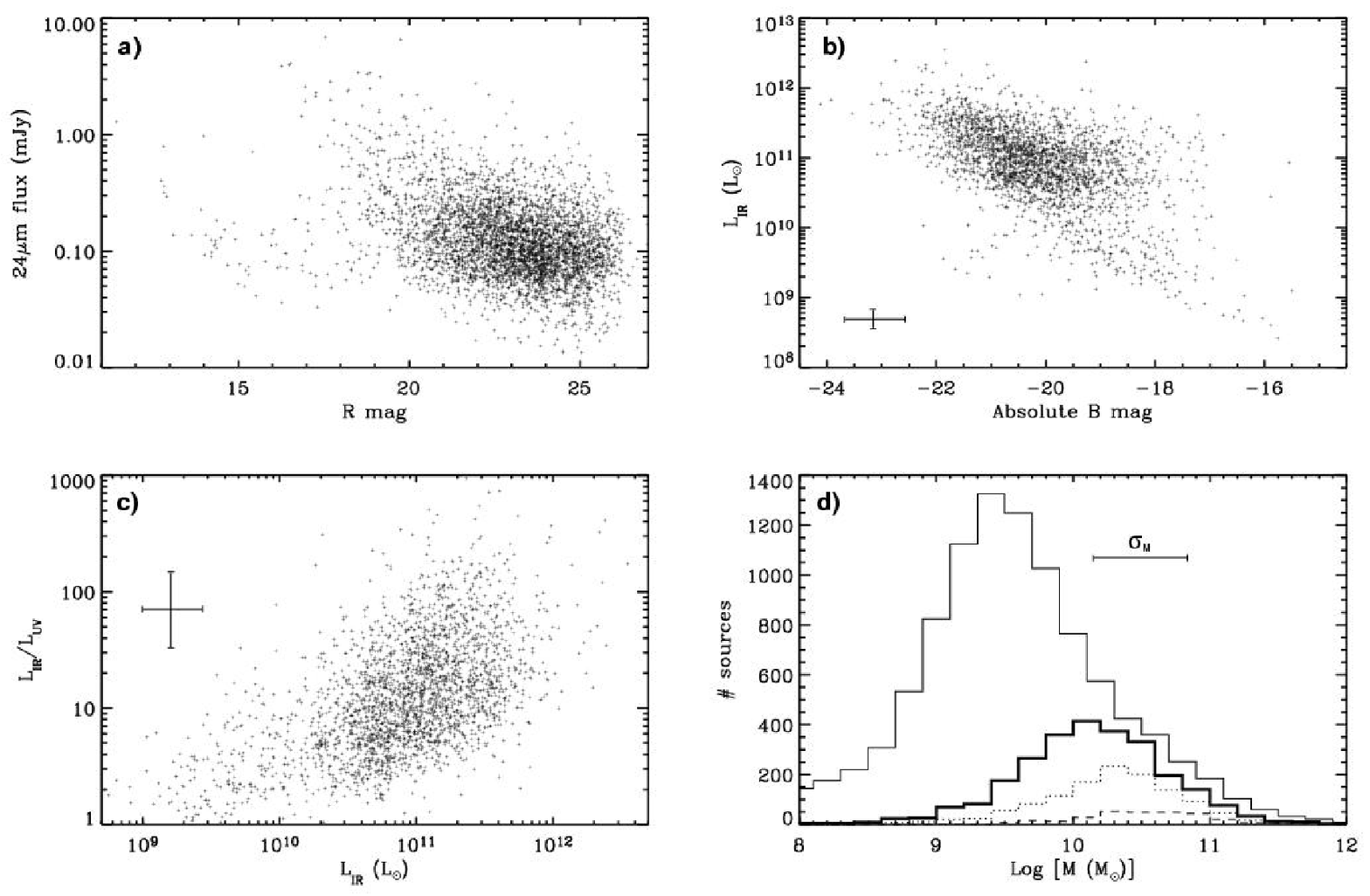}
\vskip 0.2cm \figcaption{ {\it a)\,} Observed 24\mic flux densities as a
function of the $R$-band magnitudes obtained from COMBO-17. {\it b)\,}
Total IR luminosities as derived in Sect.\,5 versus the absolute
$B$-band magnitudes.  {\it c)\,} IR to UV luminosity ratios as a
function of the total IR luminosities.
{\it d)\,} Histogram of the stellar masses (in solar units) for the sample of
MIPS sources (thick solid line) compared to the stellar masses of the COMBO-17
optically-selected galaxies (thin solid line, scaled down by a factor of 2.2).
The dotted-line (respectively, dashed-line) histogram corresponds to the
stellar mass distribution restricted to 24\mic sources with L$_{\rm IR}$\,$\geq$\,10$^{11}$\,L$_{\odot}$
(L$_{\rm IR}$\,$\geq$\,10$^{11.5}$\,L$_{\odot}$). Typical uncertainties
are indicated with error bars in panels  {\it b),}  {\it c)\,} and {\it d).}
}
\vskip .2cm
\end{figure*}

We also used the absolute magnitude estimates provided by COMBO-17 
to  derive the intrinsic luminosities of the MIPS sources
at optical wavelengths.
As an example we illustrate in Figure\,10b
 the COMBO-17 absolute magnitudes
estimated in the  $B$-band filter and corrected to our 
cosmology\footnote{The COMBO-17 catalog provides absolute quantities
assuming H$_0$\,=\,100~km~s$^{-1}$\,Mpc$^{-1}$,
$\Omega_m$\,=\,0.3 and $\Omega_{\lambda}\,=\,0.7$.}
as a function of
the total infrared luminosities
 derived in the previous section.
We observe a clear relationship between the two 
quantities, though the dispersion remains relatively
large (1$\sigma$\,$\sim$\,0.5\,dex). 
Similar relations are also obtained between the total IR luminosity
and the absolute magnitudes derived from the other broad-band filters
of COMBO-17.  They show that distant IR luminous sources (at least up
to $z$\,$\sim$\,1) are  preferentially associated with luminous optical
counterparts as in the local Universe \citep{Sanders96}.
This result confirms previous studies based on  $ISO$ surveys 
\citep{Rigopoulou02,Franceschini03,Zheng04}. It also suggests that
the dust responsible for the IR excess in such distant IR-luminous  objects
is likely distributed
within small-size regions and the corresponding reddening is not sufficient
to completely obscure the underlying galaxy.
 This interpretation is reinforced by the range of optical 
colors observed
among the MIPS 24\mic source population, very
similar to the range characterizing the local normal galaxies
 (Bell et al. 2005).

COMBO-17 also provides an estimate of the rest-frame galaxy luminosity 
at
$\lambda$=2800\AA, which can be used
to derive the ``IR excess'' in the MIPS sources. IR excess is usually
defined as the ratio between the IR and UV emission
 and it is plotted in 
 Figure\,10c 
as a function of the total IR luminosity. Owing to the tight relation
between the star-forming activity and the IR emission of galaxies
\citep{Kennicutt98}, this plot can also be read as the IR excess as a
function of an ``IR-equivalent'' star formation rate (an IR luminosity L$_{\rm IR}$\,=\,10$^{11}$\,L$_{\odot}$ 
typically corresponds to a formation of $\sim$\,17\,M$_{\odot}$ of stars per year following
standard calibrations). It illustrates the very well-known trend for
galaxies characterized by  more intense star-forming activity to be
generally dustier and more luminous at IR wavelengths
\citep[e.g.,][]{Buat02,Cardiel03,Pozzi03,Flores04}. We also see
that the fraction of UV photons not absorbed by dust in the case of
the most luminous sources (i.e., L$_{\rm
IR}$\,$\gtapp$\,10$^{11.5}$\,L$_{\odot}$) is negligible compared to
the energy reprocessed in the IR.

We finally provide an estimate of the stellar masses of the MIPS
sources based on the properties of their optical counterparts. A tight
correlation exists  between the rest-frame optical colors of
galaxies and their mass-to-light ratios
\citep{Bell01,Kauffmann03}. This relation is uncertain by a factor
$\sim$\,0.4\,dex due to combined effects of metallicity, dust, and
history of star formation in individual sources. It is however
accurate enough for the purpose illustrated here (i.e., getting a
qualitative estimate of the distribution of IR-luminous objects as a
function of mass). Following the detailed analysis and recipes by
\citet[, see their Appendix A]{Bell03}, we used the $V$-band absolute
luminosities of galaxies from COMBO-17 (converted to our cosmology)
and transformed these luminosities to stellar masses using rest-frame
$B-V$ colors. Figure\,10d illustrates the corresponding histograms for
both samples of optically-selected sources (thin solid line) and
of galaxies detected at 24\mic (thick solid line). It is very clear that
IR galaxies detected by MIPS are on average more massive (i.e.,
$\mathcal{M}$\,$\gtapp$\,10$^{9.5}$\,$\mathcal{M}_{\odot}$) than the
field population of sources selected at optical
wavelengths. Furthermore, the 24\mic objects tend to be
associated with even more massive galaxies when their
IR luminosity (and thus their star formation rate)
increases. As already observed in the case of 15\mic sources detected by $ISO$ at $z$\,$\ltapp$\,1
\citep{Franceschini03},
LIRGs with L$_{\rm IR}$\,$\gtapp$\,10$^{11.5}$\,L$_{\odot}$
have most often counterparts
with $\mathcal{M}$\,$\gtapp$\,10$^{10}$\,$\mathcal{M}_{\odot}$.
In Sect.\,8 we will discuss this issue within the more general context of the
comoving star-forming activity at $z$\,$\sim$\,1.

As a side note, we found a group of 91~sources (among which 54 have
 $f_{\rm 24\mu m}$\,$\geq$\,83\muJypa) identified with photometric
 redshifts and totally escaping the various relations illustrated on
 Figs.\,10abc.
These objects are rather faint in the optical
($R$\,$\gtapp$\,23\,mag), and their $f_{24\mu m}$/$f_{R}$ flux ratios
are particularly high and typical of LIRGs/ULIRGs. According to our
previous results, and given that most of them
have disturbed merger morphologies
 chatacteristic of high redshift star-forming
galaxies \citep[e.g.,][]{Bell05},
 they are thus likely located at
$z$\,$\gtapp$\,0.5-0.6, which is also supported by their morphologies
since many of those are chatacteristic of high redshift star-forming
galaxies.
 However they have been assigned very low
redshifts by COMBO-17, mostly in the apparent overdensity observed at
$z$\,$\sim$\,0.15 in the CDFS.  This would imply
faint infrared
luminosities
(10$^{9}$\,L$_{\odot}$\,$\ltapp$\,L$_{\rm
IR}$\,$\ltapp$\,10$^{10}$\,L$_{\odot}$, see Fig.\,9)
and  optical absolute magnitudes
much higher ($M_{r}$\,$\gtapp$\,$-15$) than expected from the correlations.
 We believe that these redshifts have probably been 
misidentified.
We decided to exclude the corresponding
sources from our sample.

\section{Infrared luminosity functions}

\subsection{Methodology}

We  explore in this section the evolution of IR
luminosity functions (LFs) at 0\,$\ltapp$\,$z$\,$\ltapp$\,1.2.
These LFs were derived for various redshift bins
 with the usual $1/V_{\rm max}$ formalism
\citep{Schmidt68,Huchra73,Felten76} applied to our sub-sample of sources
brighter than  $f_{24\mu
m}$\,$=$\,83\muJy and $R$\,=\,24\,mag. The selection function that was
used in this goal was
computed as follows. First, we considered 
the probability for a given object to verify our selection criterion
at 24\micpa. This probability equals~1 for sources brighter than
$\sim$\,300\muJy but decreases at fainter fluxes due to the
incompleteness of the 24\mic catalog. The corresponding effect was
quantified using the detailed simulations described by \citet[, see
  their figure\,1]{Papovich04}.  Second, we estimated the probability
for a 24\mic source to be associated with an optical counterpart at
$R$\,$\leq$\,24\,mag. This probability is a redshift-dependent function, and it
can be derived with an estimate of the absolute magnitude corresponding
to the optical selection criterion ($R$\,=\,24\,mag) combined with the relation
between the infrared and the optical luminosities established in the
previous section. At this step, we actually assumed that 100\% of 
sources with $R$\,$\leq$\,24\,mag are
detected by COMBO-17\footnote{This may not be true in the few regions
closely surrounding very bright stars of the optical image. However
this effect is obviously negligible over the total field of view
considered in this work.}. Finally, we took into account  the probability
for a given optical source to be characterized by a redshift. This was
done using the
estimate of the redshift identification completeness that we determined
in Sect.\,3.3
for sources at $R$\,$\leq$\,24\,mag.  

The computed selection function was subsequently used for each single galaxy
of the sample to weight the differential volume elements that are integrated
when calculating the total accessible volume  $V_{\rm
max}$. Given the multi-wavelength flux limits considered
here, this integration was performed up to the maximum
redshift enabling the detection of the object
 at {\it both\,} 24\mic {\it and\,} optical
wavelengths (i.e., the lowest value between the redshift where
the 24\mic flux would drop below 83\muJy and the redshift where
the $R$-band magnitude would reach 24\,mag).

The determination of the luminosity function was performed for each  redshift
range  independently of the sources lying in the other
bins. 
For this reason, the $V_{\rm max}$ approach is  more subject to density
fluctuations than other methods like the stepwise maximum likelihood
\cite[e.g.,][]{Efstathiou88,Willmer97}. Even though cosmic variance is not negligible in CDFS,
 we will see however  that the
uncertainties affecting our conclusions
 are largely dominated by the errors on the
$k$-corrections that are used to translate fluxes into luminosities.
 The $V_{\rm max}$
technique might also be sensitive to the
photometric redshift uncertainties, but as
 we already pointed out in
Sect.\,3.2, the effect should be negligible here given the
accuracy of the COMBO-17 redshift classification.

\subsection{Results}

From  the luminosity- dependent templates previously discussed,
we converted the 24\mic observed fluxes
into monochromatic luminosities at 12, 15, 25 and 60\mic as described in 
Sect.\,5.2. The latter were
 subsequently used to derive the  associated 
 luminosity functions that we finally compared
to the local LFs  derived from $IRAS$ and $ISO$.
In
Figure\,11 we illustrate the 15\mic LF $\psi_{15}$(L,$z$) estimated for
different redshift bins. The corresponding data points are reported in Table\,1.
At this wavelength the
$k$-corrections needed to convert  24\mic fluxes into
luminosities are minimized for most of the sources.
The majority of sources are
indeed
located at 0.5\,$\ltapp$\,$z$\,$\ltapp$\,0.8 where the MIPS 24\mic
filter  probes the 13.5--16\mic rest-frame wavength range. This behavior
reduces the dependence
of the luminosity functions
on the assumed IR templates.  We also show
in Figure\,12 the total IR LFs $\psi_{\rm IR}$(L,$z$)
computed from the infrared luminosities shown in Fig.\,9. Data points are given in 
Table\,2. These LFs are
obviously more dependent on the choice of IR SEDs since the
uncertainty in the conversion between flux and luminosity is larger
($\sim$\,0.4\,dex). They are however easier to interpret in the
context of IR galaxy evolution.

\begin{deluxetable*}{lcccccc}
\singlespace
\tablecolumns{7} 
\tablewidth{0in}
\tablecaption{ Luminosity
functions derived at 15\mic from the $1/V_{\rm max}$ analysis$^a$}
\tabletypesize{\footnotesize}
\tablehead{
\colhead{Log [L$_{15\mu m}$ (L$_\odot$)]} & \colhead{} & \multicolumn{5}{c}{$\Phi$ [\# Mpc$^{-3}$ (Log L$_{15\mu m}$)$^{-1}$]} \\
\colhead{}  & \colhead{} & \multicolumn{5}{c}{ ------------------------------------------------------------------------------------------------------------------------------------------ } \\
\colhead{} & \colhead{} & \colhead{0.3\,$\leq$\,$z$\,$\leq$\,0.45} & \colhead{0.45\,$\leq$\,$z$\,$\leq$\,0.6} & \colhead{0.6\,$\leq$\,$z$\,$\leq$\,0.8} & \colhead{0.8\,$\leq$\,$z$\,$\leq$\,1.0} & \colhead{1.0\,$\leq$\,$z$\,$\leq$\,1.2}} 
\startdata
 9.0 ...........  & &  5.80$_{-2.58}^{+2.52}$\,$\times$\,10$^{-3}$  &                                             &                                             &                                             &                                             \\
 9.5 ...........  & &  2.93$_{-1.30}^{+1.26}$\,$\times$\,10$^{-3}$  & 2.62$_{-1.16}^{+1.22}$\,$\times$\,10$^{-3}$ &                                             &                                             &                                             \\  
10.0 ...........  & &  5.44$_{-3.02}^{+3.12}$\,$\times$\,10$^{-4}$  & 1.75$_{-0.78}^{+0.86}$\,$\times$\,10$^{-3}$ & 2.90$_{-1.29}^{+1.39}$\,$\times$\,10$^{-3}$ & 2.27$_{-1.01}^{+1.13}$\,$\times$\,10$^{-3}$ & 2.38$_{-1.06}^{+1.02}$\,$\times$\,10$^{-3}$ \\
10.5 ...........  & &  8.37$_{-6.28}^{+29.6}$\,$\times$\,10$^{-5}$  & 3.78$_{-1.68}^{+1.52}$\,$\times$\,10$^{-4}$ & 7.72$_{-3.43}^{+3.56}$\,$\times$\,10$^{-4}$ & 1.40$_{-0.62}^{+0.73}$\,$\times$\,10$^{-3}$ & 1.21$_{-0.54}^{+0.67}$\,$\times$\,10$^{-3}$ \\  
11.0 ...........  & &                                               & 3.44$_{-2.58}^{+12.5}$\,$\times$\,10$^{-5}$ & 3.35$_{-2.51}^{+8.38}$\,$\times$\,10$^{-5}$ & 1.06$_{-0.67}^{+0.73}$\,$\times$\,10$^{-4}$ & 2.54$_{-1.13}^{+1.18}$\,$\times$\,10$^{-4}$ \\
11.5 ...........  & &                                               &                                             & 8.38$_{-6.29}^{+41.9}$\,$\times$\,10$^{-6}$ & 1.07$_{-0.79}^{+52.9}$\,$\times$\,10$^{-5}$ & 2.35$_{-1.76}^{+5.88}$\,$\times$\,10$^{-5}$ \\
\enddata
\tablenotetext{a}{assuming a
$\Lambda$CDM cosmology with H$_0$\,=\,70~km~s$^{-1}$\,Mpc$^{-1}$,
$\Omega_m$\,=\,0.3 and $\Omega_{\lambda}\,=\,0.7$.}
\end{deluxetable*}

\begin{deluxetable*}{lcccccc}
\singlespace
\tablecolumns{7} 
\tablewidth{0in}
\tablecaption{Total IR luminosity
functions  derived from the $1/V_{\rm max}$ analysis$^a$}
\tabletypesize{\footnotesize}
\tablehead{
\colhead{Log [L$_{\rm IR}$ (L$_\odot$)]} & \colhead{} & \multicolumn{5}{c}{$\Phi$ [\# Mpc$^{-3}$ (Log L$_{\rm IR}$)$^{-1}$]} \\
\colhead{}  & \colhead{} & \multicolumn{5}{c}{ ------------------------------------------------------------------------------------------------------------------------------------------ } \\
\colhead{} & \colhead{} & \colhead{0.3\,$\leq$\,$z$\,$\leq$\,0.45} & \colhead{0.45\,$\leq$\,$z$\,$\leq$\,0.6} & \colhead{0.6\,$\leq$\,$z$\,$\leq$\,0.8} & \colhead{0.8\,$\leq$\,$z$\,$\leq$\,1.0} & \colhead{1.0\,$\leq$\,$z$\,$\leq$\,1.2}} 
\startdata
10.0 ...........  & &  8.88$_{-5.92}^{+11.8}$\,$\times$\,10$^{-3}$  &                                             &                                             &                                             &                                             \\
10.5 ...........  & &  3.52$_{-2.35}^{+4.70}$\,$\times$\,10$^{-3}$  & 3.84$_{-2.56}^{+5.13}$\,$\times$\,10$^{-3}$ &                                             &                                             &                                             \\  
11.0 ...........  & &  0.96$_{-0.64}^{+1.28}$\,$\times$\,10$^{-3}$  & 2.29$_{-1.52}^{+3.06}$\,$\times$\,10$^{-3}$ & 3.43$_{-2.29}^{+4.58}$\,$\times$\,10$^{-3}$ & 2.77$_{-1.85}^{+3.69}$\,$\times$\,10$^{-3}$ &                                             \\
11.5 ...........  & &  0.16$_{-0.13}^{+0.42}$\,$\times$\,10$^{-3}$  & 0.53$_{-0.35}^{+0.71}$\,$\times$\,10$^{-3}$ & 1.18$_{-0.79}^{+1.57}$\,$\times$\,10$^{-3}$ & 1.78$_{-1.19}^{+2.37}$\,$\times$\,10$^{-3}$ & 1.11$_{-0.74}^{+1.48}$\,$\times$\,10$^{-3}$ \\  
12.0 ...........  & &                                               & 6.87$_{-5.15}^{+17.2}$\,$\times$\,10$^{-5}$ & 8.38$_{-6.29}^{+13.2}$\,$\times$\,10$^{-5}$ & 2.86$_{-1.91}^{+3.81}$\,$\times$\,10$^{-4}$ & 5.91$_{-3.94}^{+7.88}$\,$\times$\,10$^{-4}$ \\
12.5 ...........  & &                                               & 1.71$_{-1.29}^{+2.58}$\,$\times$\,10$^{-5}$ & 8.38$_{-6.29}^{+12.6}$\,$\times$\,10$^{-6}$ & 4.23$_{-3.17}^{+10.6}$\,$\times$\,10$^{-5}$ & 5.88$_{-4.41}^{+9.30}$\,$\times$\,10$^{-5}$ \\
13.0 ...........  & &                                               &                                             &                                             &                                             & 1.18$_{-0.88}^{+4.16}$\,$\times$\,10$^{-5}$ \\
\enddata
\tablenotetext{a}{assuming a
$\Lambda$CDM cosmology with H$_0$\,=\,70~km~s$^{-1}$\,Mpc$^{-1}$,
$\Omega_m$\,=\,0.3 and $\Omega_{\lambda}\,=\,0.7$.}
\end{deluxetable*}

\begin{figure*}[htpb]
\includegraphics*[width=18cm]{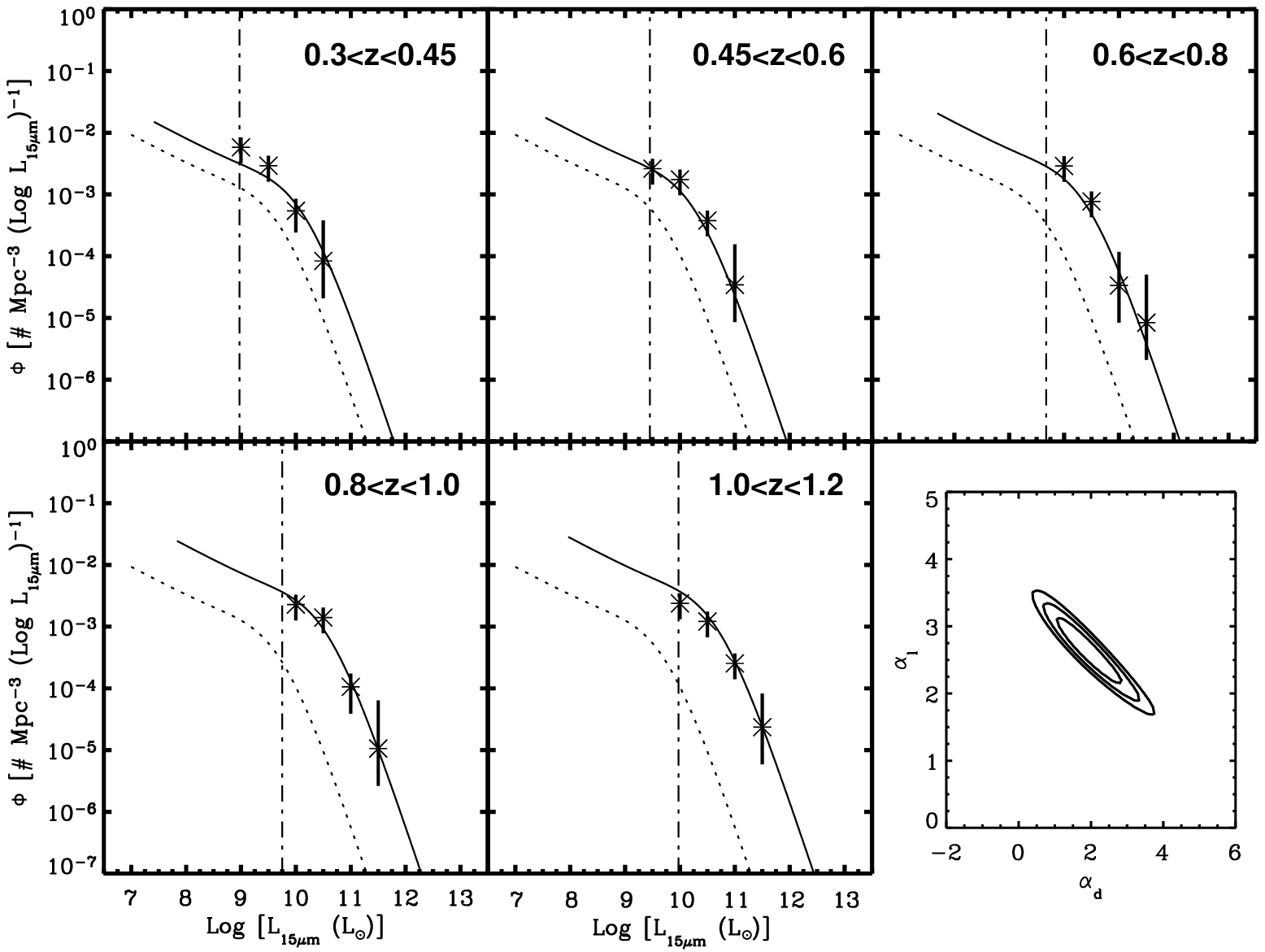}
\vskip 0.2cm\figcaption{ The 15\mic luminosity function estimated per redshift bin
between $z$\,=\,0 and $z$\,=\,1.2 ($\ast$~symbols) with the $1/V_{\rm
max}$ formalism.  The 3$\sigma$ uncertainties are indicated with
vertical solid lines (see text for more details).  Data points are
fitted by evolving the local 15\mic LF (taken from \citealt{Xu00} and
represented as a dotted line
in each panel) both
in luminosity and density ($\alpha_D$\,=\,2.1, $\alpha_L$\,=\,2.6,
solid line).
Vertical dash-dotted lines denote the luminosity corresponding
to an observed 24\mic flux of 83\muJy (i.e., 80\% completeness of the
MIPS survey) calculated at the lowest redshift considered in each panel.
The inset represents the 1$\sigma$, 2$\sigma$ and 3$\sigma$ iso-probability
contours of the likelihood estimated as a function of 
$\alpha_D$ and $\alpha_L$ with a $\chi^2$ test.}
\vskip .2cm
\end{figure*}

\begin{figure*}[htpb]
\includegraphics*[width=18cm]{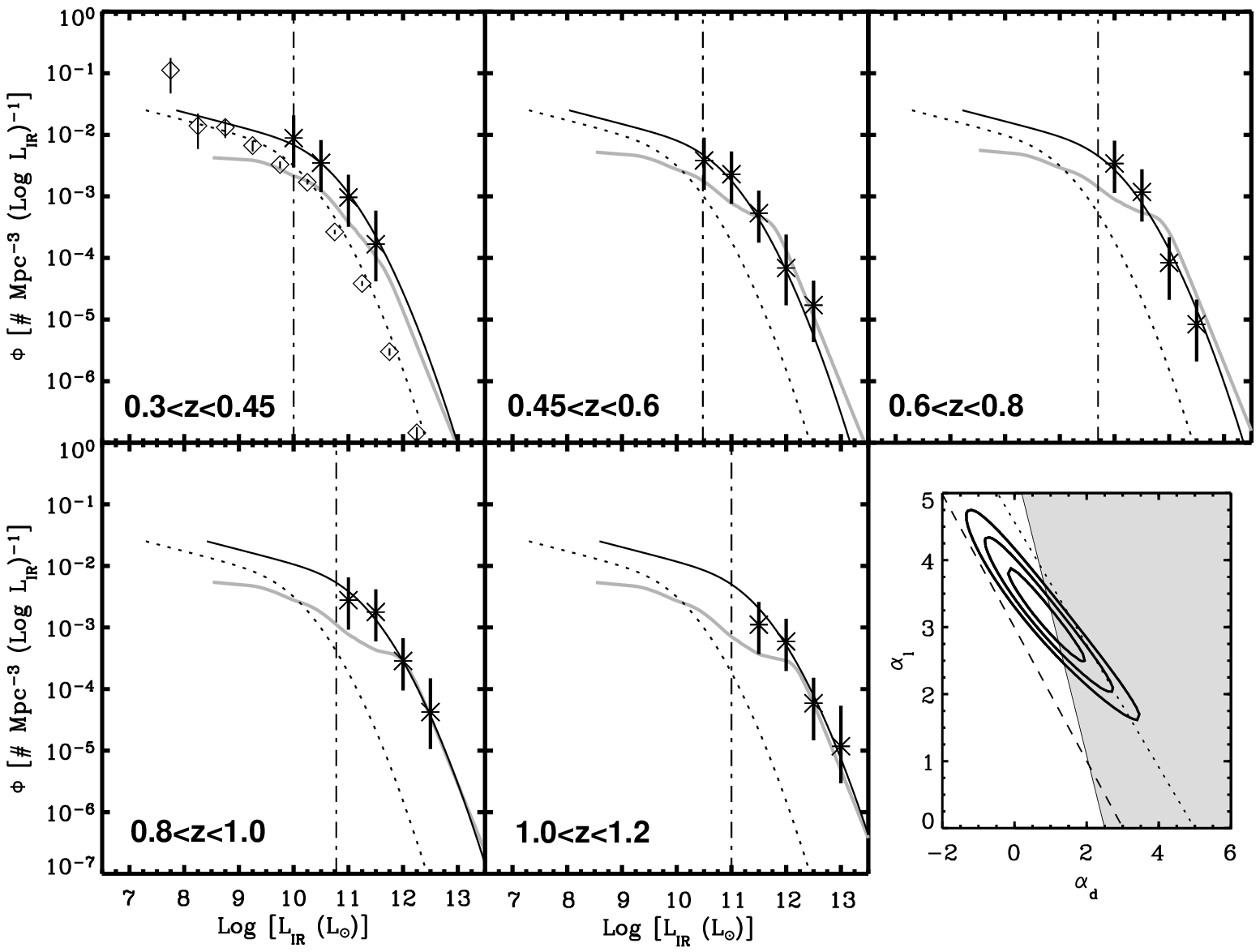}
\vskip 0.2cm \figcaption{Evolution of the total IR luminosity function
(same legend as in Figure\,11).  The fit has been obtained by
evolving
the total IR local LF (shown as a dotted line in each panel)
with $\alpha_D$\,=\,1.0 and $\alpha_L$\,=\,3.15.  This IR local
luminosity function was estimated from the 60\mic LF of \citet[, see
also Saunders et al. 1990]{Takeuchi03} and the 60\micpa/total-IR IRAS
correlation \citep[e.g.,][]{Chary01}. It is compared on the first
panel with the total-IR LF derived from the IRAS revised Bright Galaxy
Sample \citep[open diamonds,][]{Sanders03}.  The solid grey line
represents the total IR luminosity function from the model by
\citet{Lagache04}. In the inset, the
 dotted and dashed lines delimit a region where the parameters
$\alpha_D$ and $\alpha_L$ lead to an evolution of the star formation
activity that is consistent with the constraints determined by
\citet[, see text for details]{Hopkins04}.  The shaded region
corresponds to the excluded parameter space that leads to an
overproduction of the 24\mic counts at faint fluxes. This constraint
reduces significantly the observed degeneracy.}
\vskip .2cm
\end{figure*}

To correct for cosmic variance effects in each redshift bin, the
 luminosity functions were normalized by the ratio between the
 $B$-band luminosity densities produced by galaxies of the blue
 sequence and measured respectively in the CDFS and over the 3 fields
 of COMBO-17 \citep{Wolf03,Wolf04,Bell04}.  Even though the $B$-band
 and the IR selected populations  differ in their evolution with
 redshift, they both trace instantaneous star-forming activity (yet
 with different time scales). This connection results in an obvious relation
 between the two that we already highlighted in Sects.\,3.4 \&~6 (see
 Fig.\,10b, see also Bell et al. 2005). 
The characterization of the $B$-band luminosity
 function over different fields can therefore be used to estimate a
 first-order correction and minimize the cosmic variance affecting our
 24\mic sample. With its redshift peak at $z$\,$\sim$\,0.65, the CDFS
 is particularly subject to this effect.

\subsection{Uncertainties}

For each luminosity bin in a given redshift range, uncertainties
$\sigma(L_i,z_j)$ were estimated as the combination of (i) poisson
noise statistics on the number of sources used in the measurement
(i.e., the rms error $\sqrt{\Sigma V_{\rm max,k}^{-2}}$, with ``k''
the source index in the bin) and (ii) uncertainties in the flux density
  at 24\mic and the conversion into luminosities. The effect of the
  latter was simulated with a Monte Carlo approach.  We assigned to
  each galaxy a range of luminosities (monochromatic or integrated)
  that were calculated by taking account of the nominal flux and
  uncertainty at 24\mic as well as the different possible SEDs for the
  conversion.  We found this conversion from flux to luminosity to be
  by far the dominant source of uncertainty, which explains the larger
  error bars found in the case of $\psi_{\rm IR}$ than for
  $\psi_{15}$.
Regarding the monochromatic
15\mic LF however, flux uncertainties at 24\mic also contribute
significantly, especially at the faintest levels.  For each redshift
bin finally, we uniformly added another uncertainty (0.15\,dex) to the
whole luminosity function based on our estimate of the cosmic variance
effect. Because it should be negligible, we did not simulate the effect related to photometric redshift
uncertainties. 

\subsection{Parameterization and evolution with redshift}

In both Figures\,11 \& 12 we have also illustrated
the luminosity functions determined locally from $IRAS$ and $ISO$.
The 15\mic local LF $\phi_{\rm 15}(L)$ was taken from 
\citet{Xu98} and \citet{Xu00}. Regarding the total
IR luminosity function $\phi_{\rm IR}(L)$ at $z$\,=\,0, we show on one hand 
a recent result from the revised IRAS Bright Galaxy Sample \citep{Sanders03}, on the
other hand an estimate that we
 derived
from the 60\mic local LF \citep{Saunders90,Takeuchi03} assuming the
tight correlation between the 60\mic and the total IR luminosity of
galaxies. Not surprisingly, these two estimates agree well with each other.
A commonly known characteristic of
these IR luminosity functions is
their relatively large number of sources at the
bright end. As a result they are better described
 by a double-exponential
profile rather than a  classical
Schechter parameterization. Their usual analytical form  is given as
follows:

\begin{eqnarray}
\phi_\lambda(L)&=&\frac{dN(L)}{\;dV\;d{\log}_{10}(L)}\\
       &=&\phi^{\star}_\lambda\left(\frac{L}{L^{\star}_\lambda}\right)^{1-\alpha_\lambda}\exp\left[-\frac{1}{2{\sigma_\lambda}^2}{\log}^2_{10}\left(1+\left(\frac{L}{L^{\star}_\lambda}\right)\right)\right]\nonumber
\end{eqnarray}

\noindent where $dV$ is the differential element of comoving volume  and $dN(L)$ the number of sources with
a luminosity $L$ within $dV$ and per bin of $d$log$_{10}(L)$.

~

As expected, the comparison between
 $\psi_{\lambda}$(L,$z$)  and  the local luminosity functions
reveals 
a very strong evolution of the LFs with
lookback time.
 For each
redshift bin,  $\psi_{\lambda}$(L,$z$) was fitted assuming a
monotonic evolution of the local LF $\phi_{\lambda}(L)$ in both luminosity and
density as a function of (1+$z$):

\begin{equation}
\psi_{\lambda}(L,z)=g(z)\phi_{\lambda}\left(L/f(z)\right)
\end{equation}

\noindent with $g(z)=(1+z)^{\alpha_{\rm D}}$ and $f(z)=(1+z)^{\alpha_{\rm L}}$.

~

We explored the possible range of values for the fitting parameters
$\alpha_{\rm D}$ and $\alpha_{\rm L}$ using a $\chi^2$
minimization. This $\chi^2$ was computed from the difference between
the observed and the fitted luminosity functions over the
0.3\,$\leq$\,$z$\,$\leq$\,1.0 redshift range:

\begin{equation}
\chi^{2}_\lambda \propto \sum_{L_i} \sum_{0.3\leq z_j \leq 1} \left[\frac{\psi_{\lambda,obs.}-\psi_{\lambda,model}}{\sigma} \right]^{2}_{L_i,z_j}
\end{equation}

~ 

Because of  their lower statistical significance, LFs determined at 
 $z$\,$\leq$\,0.3 and $z$\,$\geq$\,1
were  ignored in this process. Reduced $\chi^2$ values
were finally transformed into
 likelihood estimates using log$(L)=-0.5\chi^2$.
The corresponding 1$\sigma$, 2$\sigma$ and 3$\sigma$ iso-probability
contours for $\psi_{15}$ and $\psi_{\rm IR}$ are displayed in the
insets of Figures\,11 \&~12 respectively.

The elliptical shape and the orientation of these contours in the
luminosity and density evolution
parameter space
reflect a   well-known
degeneracy often encountered when fitting high redshift luminosity functions.
Given the sensitivity of our 24\mic survey indeed,  
the very faint-end portion of $\psi_{\lambda}$(L,$z$) cannot be 
constrained at any redshift.
The LF can thus be described with a variety of scenarios combining
different amount of evolution in $\alpha_{\rm D}$ and
$\alpha_{\rm L}$. The constraints that we obtained can be summarized
 as follows:

\begin{eqnarray*}
\psi_{15}(L,z):   
\begin{cases}
\alpha_L = 2.6 \pm 0.9 ~ (3\sigma) \\
\alpha_D = (-1.80 \times \alpha_L + 6.75) \pm 1.1 
\end{cases}
\end{eqnarray*}
\begin{eqnarray*}
\psi_{IR}(L,z):   
\begin{cases}
\alpha_L = 3.15 \pm 1.6 ~ (3\sigma) \\
\alpha_D = (-1.55 \times \alpha_L + 5.90) \pm 1.6 
\end{cases}
\end{eqnarray*}

~

~

We will analyze in Sect.\,7.4 how this degeneracy can be broken by
taking account of other independent constraints.  Nonetheless we
stress that the knee of the luminosity functions is well detected
throughout the redshift range considered in this work (i.e.,
$z$\,$\ltapp$\,1).  This allows us to exclude a pure evolution of the
LF in density (i.e., $\alpha_{\rm L}$\,=\,0) with a very high
confidence. 

Because of the quasi-linearity between  L$_{15}$ and L$_{\rm IR}$
\citep{Chary01,Takeuchi05}, these constraints on the evolution
of  $\psi_{15}$ and $\psi_{\rm IR}$ should be in principle strictly
similar. While they do agree rather well with each other,
the extension of the contours reveal however that the evolution of
the 15\mic LF
seems to be better constrained than the evolution of  $\psi_{\rm IR}$.
As we already pointed out, LF uncertainties are
largely dominated by the errors in the conversion between the 24\mic
flux and the luminosities,  and these errors  are in fact minimized around
15\micpa. Furthermore, we see that $\psi_{15}$
 might be   characterized  by a 
stronger evolution in density and a smaller evolution
in luminosity compared to  $\psi_{\rm IR}$.
This discrepancy could be a consequence of the non-negligible
dispersion that exists in the correlations observed between fluxes at
mid- and far-infrared wavelengths \citep[e.g.,][]{Helou86,Xu00,Dale01,Dale02}.
 Indeed the templates that we used
to extrapolate fluxes to luminosities do not really show a pure linear
relation between L$_{15}$ and L$_{\rm IR}$ (see Fig.\,8).  Moreover, the 
constraints on $\alpha_D$ and $\alpha_L$ strongly depend on the
assumed local luminosity functions, and the comparison between
$\phi_{15}$ and $\phi_{60}$ (that we used to compute $\phi_{\rm IR}$)
clearly reveals a non-linearity between the two. Finally, it might be
 suggested that 
our scenario based on 
 a single evolving population is overly simplistic.
Considering distinct object types
characterized by their own evolution
 \citep{Xu00,Lagache03,Lagache04,Pozzi04,Gruppioni05},
 bi-variate LFs \citep{Chapman03a,Lewis05} or luminosity functions
  only evolving at
their bright end \citep{Chary01} could be one way to explore this 
effect in more detail.

In spite of this apparent degeneracy, it should be however
noted that the evolution of
  the total luminosity density $\Omega_{\rm IR}$
integrated from these infrared LFs is
  more tightly constrained than the evolution of the characteristic parameters
$L^{\star}_{\lambda}$ and $\phi^{\star}_{\lambda}$
 considered separately.  In
  each redshift bin, MIPS can indeed detect those sources responsible for the bulk of the
  comoving luminosity density, and the
  additional uncertainty related to the extrapolation for
  taking  account of the contribution of faint objects (i.e., faint-end slope of the LF, see Sect.\,7.5)
is therefore not dominant. 
Based on the relation that we obtained
between $\alpha_{\rm L}$ and $\alpha_{\rm D}$, we find that $\Omega_{\rm IR}$
evolves as (1+$z$)$^{3.9\pm0.4}$ at 0\,$\ltapp$\,$z$\,$\ltapp$\,1 (see also 
Sect.\,8 and Fig.\,14).

Finally,  we tested the effect of our redshift identification incompleteness
at $z$\,$\gtapp$\,1 on the LF estimates by
computing
$\chi^2$ including the observed luminosity function determined at
1\,$\leq$\,$z$\,$\leq$\,1.2. For both $\psi_{15}$ and $\psi_{\rm IR}$
the minimization of $\chi^2$ in this case
leads to a short translation of the iso-probability contours toward
smaller values of $\alpha_D$ and larger values of $\alpha_L$
compared to what we previously found. 
This trend is consistent with expectations if $\psi_\lambda$
is underestimated at $z$\,$\gtapp$\,1. Indeed the latter 
artificially yields
a fainter evolution in density, the effect of
which gets compensated in the lowest redshift bins thanks to a slight
increase in luminosity. 
As a
sanity check we finally computed  $\chi^2$ also excluding the LF
obtained at 0.8\,$\leq$\,$z$\,$\leq$\,1 where we may also be missing a
few 24\mic sources fainter than $R$\,$\sim$\,24\,mag. The results that
we found remain  consistent with those initially obtained.

\subsection{Breaking degeneracies}

\citet{Hopkins04} combined constraints from source
number counts at radio wavelengths with estimates of
the comoving star formation rate (SFR) density of the Universe at high
redshift to break the degeneracy that also arises when quantifying
the evolution of
 star-forming radio-selected galaxies.
Following his method we investigate
in this section how similar considerations can help in better
constraining the evolution of IR galaxies.

Up to $z$\,$\sim$\,1, the integrated star formation density per
comoving volume of the Universe $\dot{\rho}_\star(z)$ is now well
established within a factor of 2 to 3 \citep{Hopkins04}. Given the
relationship between the total IR emission of galaxies and their obscured SFR
 \citep{Kennicutt98}, the evolution of $\psi_{\rm IR}$ 
 converted into a history of the total dusty star-forming activity
  must 
be therefore consistent with these constraints on $\dot{\rho}_\star(z)$.
We computed the ``IR-equivalent  SFR''
as a function of redshift with different combinations of $\alpha_D$ and $\alpha_L$. The
dispersion in the relation between the flux at 24\mic and the total IR
luminosity 
obviously affects this estimate.  Moreover we stress that it should
 only be a lower limit on the true SFR given the un-absorbed UV
 photons produced by young stars and not accounted for by the IR
 measurements \citep[e.g.,][, see also the L$_{\rm IR}$/L$_{\rm UV}$
 ratio of the MIPS sources in Fig.\,10c]{Bell03a}.  Given these
 caveats and the additional uncertainty on $\dot{\rho}_\star(z)$
 mentioned above,
we required our IR-SFR determination to lie within 0.5\,dex of the averaged relation
between log$_{\rm 10}$($\dot{\rho}_\star(z)$) and log$_{\rm
10}$(1+$z$) derived by \citet{Hopkins04}.  This constraint demarcates
a specific region in the luminosity and density evolution parameter
space, lying between the dashed and the dotted lines shown in the inset of
Figure\,12.  As we can see, the evolving parameters $\alpha_D$ and $\alpha_L$
 that we derived
from  $\chi^2$ minimization in the previous section agree well
with the known history of star formation up to $z$\,$\sim$\,1.
However we also note that this approach does not really help in solving the aforementioned
degeneracy. One may need a more accurate determination of
$\dot{\rho}_\star(z)$ to progress in this direction. Alternatively,
a better constraint on the conversion between $f_{24\mu m}$, L$_{\rm
  IR}$ and SFR might provide in the future a more accurate
determination of $\dot{\rho}_\star(z)$.

A much  more interesting constraint can be obtained from the source number counts.
In fact we only considered 24\mic sources brighter than 83\muJy when building the luminosity
functions $\psi_\lambda(L,z)$, but the counts at lower fluxes
can be used to derive limits on the contribution of sources at
luminosities fainter than those taken to minimize the $\chi^2$. Using
the 24\mic observations of the ``GOODS test-field'' centered on
ELAIS-N1, \citet{Chary04} and \citet{Papovich04} constrained the faint
source density down to $f_{24\mu m}$\,$\sim$\,30\muJy and showed that
the differential counts normalized to the Euclidian slope are
likely dropping very rapidly below this limit.
We must therefore ensure that the evolution of the infrared LF
does not lead
to an overproduction of these counts at faint fluxes.
To check the latter
 we derived as a function of $\alpha_D$ and $\alpha_L$ the differential number
counts that would be produced up to $z$\,=\,1 by a population of sources
described by $\psi_\lambda(L,z)$ and evolving as given by Equation\,2
(see Fig.\,13).
 We rejected the solutions overproducing the total counts obtained
from the GOODS test field at the faint end.  Surprisingly we found
that this constraint is nearly independent of the library of IR SEDs
used to describe the galaxy population.
This excludes a region of the [$\alpha_D$,$\alpha_L$] parameter space that
is illustrated by the shaded area in the inset of Figure\,12.

This additional constraint  reduces significantly
the number of possibilities to describe the evolution of
$\psi_{\rm IR}(L,z)$. 
The best parameters  quantifying this evolution at 0\,$\ltapp$\,$z$\,$\ltapp$\,1 are given by
 $L^\star_{\rm IR} \propto (1+z)^{3.2_{-0.2}^{+0.7}}$  and
$\phi^\star_{\rm IR} \propto (1+z)^{0.7_{-0.6}^{+0.2}}$ (quoted uncertainty of
1$\sigma$) . They are summarized in Table\,3. 
 In particular they exclude any solution favoring
a larger evolution in density than in luminosity.
 This trend
has already
been reported by several groups using similar analysis with the $ISO$
and
SCUBA number counts,
and the overproduction of the 
background resulting from a too strong increase of $\phi^\star_{\rm IR}$
is a well-known constraint on the 
backward evolutionary scenarios of IR galaxies. It has also been noted
with models assuming pure density evolution in the Press Schechter
formalism \citep{Mould03}.

\begin{minipage}[b]{8.5cm}
\vskip .3cm 
\centerline{\psfig{file=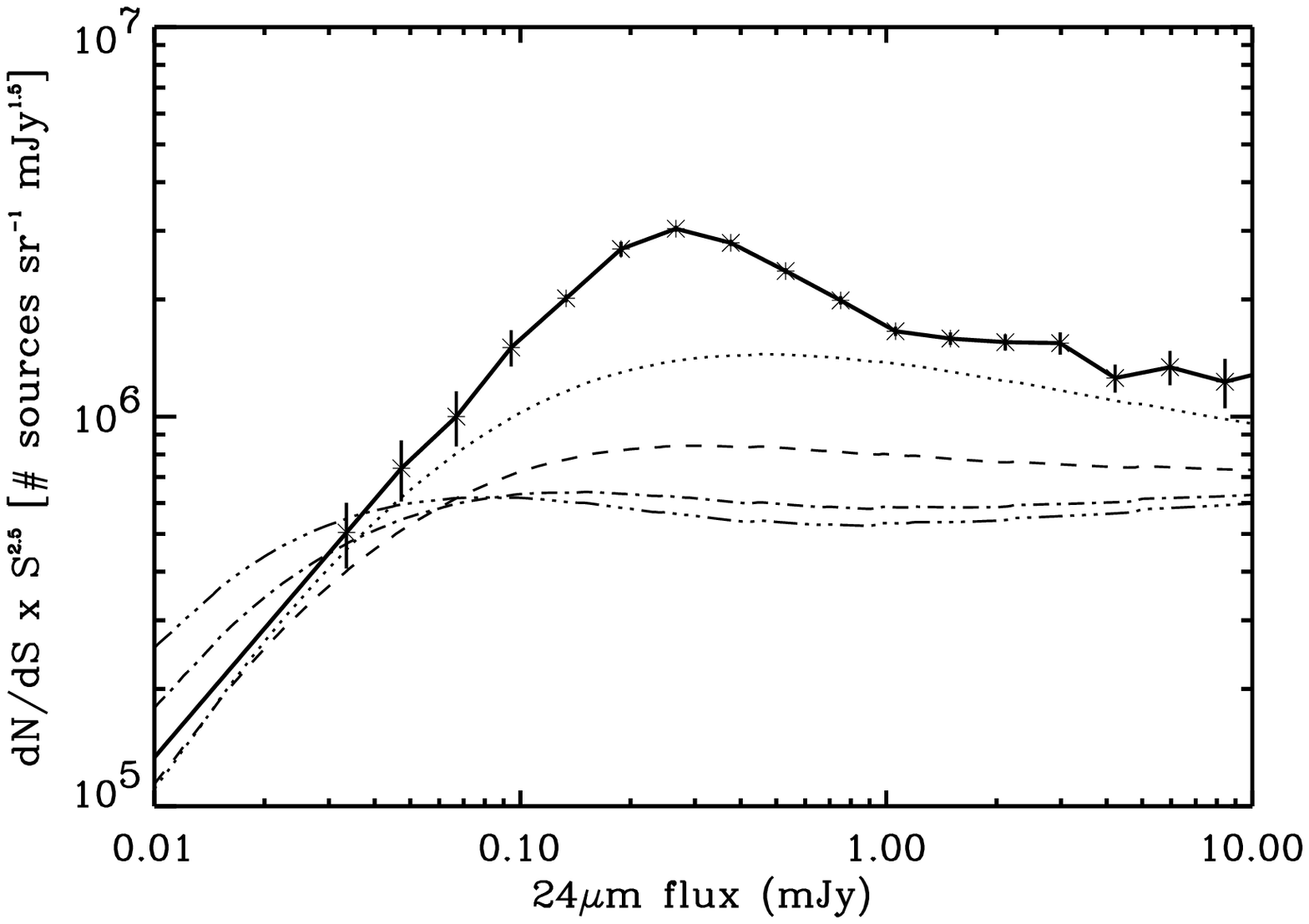,width=8.5cm,angle=0}}
\vskip 0.2cm \figcaption{Simulated 24\mic number counts produced
up to $z$\,=\,1 by a population tied to the local IR luminosity
function and evolving in luminosity and density according to various
scenarios (dotted line: $\alpha_D$=0.0, $\alpha_L$=4.0 --
dashed line: $\alpha_D$=0.2, $\alpha_L$=3.25 -- 
dash-dotted line: $\alpha_D$=1.5,  $\alpha_L$=2.75 --
triple dot-dashed line: $\alpha_D$=2.5,  $\alpha_L$=2.2).
They are compared to the total observed 24\mic source 
number counts ('$\ast$' symbols, solid line and vertical error bars) 
constrained and extrapolated at faint fluxes using the ``GOODS test field''
\citep{Chary04,Papovich04}. Note that a too large
evolution in density clearly overproduces the counts at the faint
end.}
\vskip .2cm
\label{fig:z_distrib}
\end{minipage}

\begin{deluxetable*}{lrcrccc}
\singlespace
\tablecolumns{4} 
\tablewidth{0in}
\tablecaption{Parameterization of the evolution of the total IR luminosity
function as a function of redshift$^a$}
\tabletypesize{\footnotesize}
\tablehead{
\colhead{Redshift range} & \colhead{$\phi^{\star}_{\rm IR}$} & \colhead{~} & \colhead{$L^{\star}_{\rm IR}$} & \colhead{~} & \colhead{$\alpha_{\rm IR}$} & \colhead{$\sigma_{\rm IR}$}}  

\startdata
$z$\,=\,0 (local IRAS luminosity function)      &                   0.89\,$\times$\,10$^{-2}$ &  &                    1.77\,$\times$\,10$^{9}$ &  & 1.23 & 0.72\\
0.0\,$\leq$\,$z$\,$\leq$\,0.3 ................. & 0.98$^{+0.03}_{-0.08}$\,$\times$\,10$^{-2}$ & ~&  2.77$^{+0.28}_{-0.08}$\,$\times$\,10$^{9}$ &~ & 1.23 & 0.72\\
0.3\,$\leq$\,$z$\,$\leq$\,0.45 ................ & 1.11$^{+0.07}_{-0.19}$\,$\times$\,10$^{-2}$ &  &  4.91$^{+1.23}_{-0.30}$\,$\times$\,10$^{9}$ &  & 1.23 & 0.72\\
0.45\,$\leq$\,$z$\,$\leq$\,0.6 ................ & 1.20$^{+0.11}_{-0.27}$\,$\times$\,10$^{-2}$ &  &  6.84$^{+2.35}_{-0.55}$\,$\times$\,10$^{9}$ &  & 1.23 & 0.72\\
0.6\,$\leq$\,$z$\,$\leq$\,0.8 ................. & 1.29$^{+0.14}_{-0.35}$\,$\times$\,10$^{-2}$ &  &  9.7$^{+4.4}_{-1.0}$\,$\times$\,10$^{9}$    &  & 1.23 & 0.72\\
0.8\,$\leq$\,$z$\,$\leq$\,1.0 ................. & 1.40$^{+0.19}_{-0.45}$\,$\times$\,10$^{-2}$ &  & 13.8$^{+7.8}_{-1.7}$\,$\times$\,10$^{9}$    &  & 1.23 & 0.72\\
1.0\,$\leq$\,$z$\,$\leq$\,1.2 ................. & 1.50$^{+0.24}_{-0.54}$\,$\times$\,10$^{-2}$ &  & 19.0$^{+13.0}_{-2.6}$\,$\times$\,10$^{9}$   &  & 1.23 & 0.72\\
\enddata
\tablenotetext{a}{assuming a
$\Lambda$CDM cosmology with H$_0$\,=\,70~km~s$^{-1}$\,Mpc$^{-1}$,
$\Omega_m$\,=\,0.3 and $\Omega_{\lambda}\,=\,0.7$.}
\end{deluxetable*}

Obviously we could also use this method to further restrict the
possible combinations of $\alpha_D$ and $\alpha_L$ by rejecting
scenarios that overproduce the counts at bright fluxes (i.e.,
$f_{24\mu m}$\,$\gtapp$\,0.3\,mJy) or do not reproduce the cumulative
counts discussed in Section\,4.  For instance the model predictions
represented by the dashed line in Fig.\,13 are consistent with the
constraints imposed by the faint source counts but severely
underestimate the true contribution of sources at $z$\,$<$\,1.
However we found that reproducing the bump observed at $f_{24\mu
m}$\,$\sim$\,0.3\,mJy is very dependent on the choice of SEDs, which could
also explain why the pre-launch models had
over-predicted the MIPS 24\mic number counts at these bright fluxes.
Since our goal is to exclude
unphysical solutions without being subject to the assumed SEDs, we did
not consider further this information.

\section{Discussion}

\subsection{Evolution and shape of the infrared luminosity function}

In the previous section we have constrained the knee and the bright
end of the infrared luminosity function $\psi_{\rm IR}(L,z)$ up to
$z$\,$\sim$\,1. As expected our results indicate a very strong
evolution of this LF with lookback time. We find that the space density
of galaxies with L$_{\rm IR}$\,$\geq$\,10$^{11}$\,L$_\odot$  at $z$\,$\sim$\,1
exceeds by more than 100 their density in the local Universe, which is
in fairly good agreement with the results from $ISO$.
The quantification of the evolution depends however {\it very
strongly\,} on the mid- and far-IR 
SEDs  used to compute the $k$-corrections.
This points to an urgent need for a more accurate characterization 
of infrared spectral energy distributions of normal and
luminous galaxies. This goal  might
be achieved by combining 24, 70 and 160\mic broad-band MIPS imaging
with infrared spectroscopy from the IRS spectrograph and the MIPS ``SED
mode''.

Given the limitations due to the uncertainties of our LF estimates, we
do not find any 
evidence for a modification of the shape of $\psi_{\rm IR}$
relative to the local luminosity function
at least in the luminosity range probed with MIPS.
Obviously the density of  sources fainter than the 80\%
completeness limit of the 24\mic survey
 is not directly constrained and we cannot exclude a slight
steepening of the LF faint-end slope
  at high redshifts.
However,  the weight of  this LF by  luminosity
 shows that the increase of energy produced by faint
 objects in the case of  a steeper slope would be
 mostly driven by
sources just below the knee of  $\psi_{\rm IR}$,
the faintest ones having a negligible contribution 
in spite of their larger number. 
The corresponding
effect is then very similar to the one
produced by a too large evolution of $\psi_{\rm IR}(L,z)$ in density.
As we saw in the previous section it would result in an overproduction
of the number counts at faint fluxes and violate therefore the
constraints on the background.  While
the steepening of $\psi_{\rm IR}$ at the faintest end cannot be
definitely ruled out, it must have marginal significance and will not
affect our further discussion. 

Furthermore we do not notice any obvious break in the shape of the
luminosity function $\psi_{\rm IR}(L,z)$.  Contrary to our
simple scenario that considers the MIPS detections as a single class
of objects,
these breaks can occur when the LF is
decomposed into the contribution of several populations 
(e.g., starbursts, AGNs, cold galaxies,~...)
evolving
independently with the redshift. As an example, they can be observed in the
model proposed by \citet{Lagache03,Lagache04}.
Based on
a simple decomposition of high redshift galaxies into normal non-evolving
sources and starbursts undergoing strong evolution, this model
had successfully reproduced previous IR/sub-mm
observations from $ISO$, SCUBA and COBE, but
it fails in explaining
the faint-end part of our luminosity functions because of 
a predicted break that is not observed (see Fig.\,12).
{\it Spitzer\,}  provides therefore new constraints on this kind of scenario,
and our results  suggest in this case a smooth transition between
populations so that  continuity is still observed in the 
total luminosity function. This effect could be taken into account
considering multi-variate luminosity functions \citep{Chapman03a,Lewis05}.

\subsection{Comparison with results from the previous
long-wavelength surveys}

Over the last decade cosmological surveys conducted with $ISO$
mainly at 15\micpa, 90\mic and 170\mic provided  direct evidence for
 the importance of infrared luminous sources at 
0\,$\ltapp$\,$z$\,$\ltapp$\,1, while submillimeter observations
with SCUBA revealed a very high density of dusty galaxies with 
L$_{\rm IR}$\,$\geq$\,10$^{12}$\,L$_\odot$ at very high redshifts
(i.e., $z$\,$\gtapp$\,2). 
Number counts and
redshift distributions that were derived from these surveys
as well as the direct measurement
of the far-infrared background by COBE already allowed determination of some
constraints on the evolution of the infrared energy density with
redshift \citep[e.g.,][]{Blain99,Franceschini01}.

MIPS is however the first infrared instrument with good enough
 sensitivity to obtain a direct measurement of the mid-IR luminosity
 function up to $z$\,$\sim$\,1 \citep[but see][ for LFs at 
 intermediate redshifts]{Pozzi04,Serjeant04}.  It might therefore be
 worth comparing the constraints obtained from this work with earlier
 results from the literature.
First, our 24\mic survey not only excludes very clearly the
possibility of a pure evolution of the IR luminosity function in
$\phi^\star$, but it also reveals a much stronger evolution of $\psi_{\rm
IR}$ in luminosity than in density (i.e., $\alpha_L\gg\alpha_D$). This
trend confirms previous interpretations of long-wavelength surveys and
implies that the contribution of dusty luminous galaxies 
 was significantly more important in the past.  Our evolution
constraints (i.e., $L^\star_{\rm IR}$\,$\propto$\,(1+$z$)$^{\sim 3.5}$
and $\phi^\star_{\rm IR}$\,$\propto$\,(1+$z$)$^{\sim 0.5}$) show 
good agreement with previously published analysis
\citep[e.g.,][]{Blain99b,Blain99,Franceschini01,Lagache03,Chapman02,Lewis05}.
However our data do not seem to be consistent with a luminosity
evolution stronger than $\alpha_L \sim 4.5$ as in several scenarios
examined by \citet{Chary01} and \citet{Xu00}, unless one adds a 
           decrease in density ($\alpha_D<0$).

Our analysis moreover agrees with the constraints on the LFs of
star-forming galaxies recently 
derived from the radio sub-mJy source number counts and the history of star
formation.
An evolution  slightly lower than our estimate
  ($\alpha_L=2.7\pm0.6, \alpha_D=0.15\pm0.6$)
has been obtained with this approach by \citet{Hopkins04}, but it is still 
consistent with our results within the large uncertainties.
The small discrepancy could also originate from the
likely different levels of AGN contamination characterizing 
populations selected at  24\mic and at radio wavelengths.

While our results  on the
mid- and total IR luminosity functions are still consistent with
each other, we finally 
note that our constraints on 
 $\psi_{15}$ seem to require a stronger evolution in density and a
smaller evolution in luminosity compared to $\psi_{\rm IR}$. We have
not yet understood whether this trend is physically real and reveals the need
for a more complex
evolutionary scenario than the one  we have assumed,
or whether it is artificially produced by
the infrared SEDs  used to compute
luminosities. Interestingly, the need for a strong evolution of
the mid-IR LF in density has also been  suggested by \citet{Pozzi04}
based on the ELAIS 15\mic surveys \citep[see also][]{Gruppioni05}. 
A direct comparison between our analysis and their results 
must however be done with caution.
\citet{Pozzi04} decomposed 
 their sample  into the contribution of
non-evolving galaxies that  dominate the 15\mic counts
 at low redshift, and
 distant starbursts  responsible for the increase of the IR energy
density.
  Given the negligible role of these
starbursts
 at $z$\,$\sim$\,0, their evolution 
must be
therefore stronger than that of the global sample.
 This may explain why the
evolving parameters that they derived are slightly larger than
ours (i.e., $\alpha_L=3.5^{+1.0}_{-3.5},
\alpha_D=3.8^{+2.0}_{-2.0}$). 
Note that this approach was also used by \citet{Xu01} and led to a
qualitatively similar trend.

\subsection{The evolving contribution of infrared luminous galaxies to 
star formation activity  up to $z$\,$\sim$\,1}

The constraints on the evolution of $\psi_{\rm IR}(L,z)$ can be used
to derive the relative importance of galaxies in a given
luminosity range and how their contribution to the star formation history
evolves with redshift. Using the fit and the parameterization discussed
in the previous section we compare in Figure\,14 the evolution of the IR
energy density $\Omega_{\rm IR}$ produced by infrared luminous
galaxies (i.e., LIRGs\,$+$\,ULIRGs) and fainter sources (i.e.,
L$_{\rm IR}$\,$<$\,10$^{11}$\,L$_\odot$).
Uncertainties affecting our results
are still significant due to the degeneracy previously
described.  Accordingly the range of possible solutions was estimated
from the 3$\sigma$ iso-probability contours showed in Figure\,12
excluding the combinations of $\alpha_D$ and $\alpha_L$ that 
overproduce the counts at faint fluxes.  

\begin{figure*}[htpb]
\includegraphics*[width=18cm]{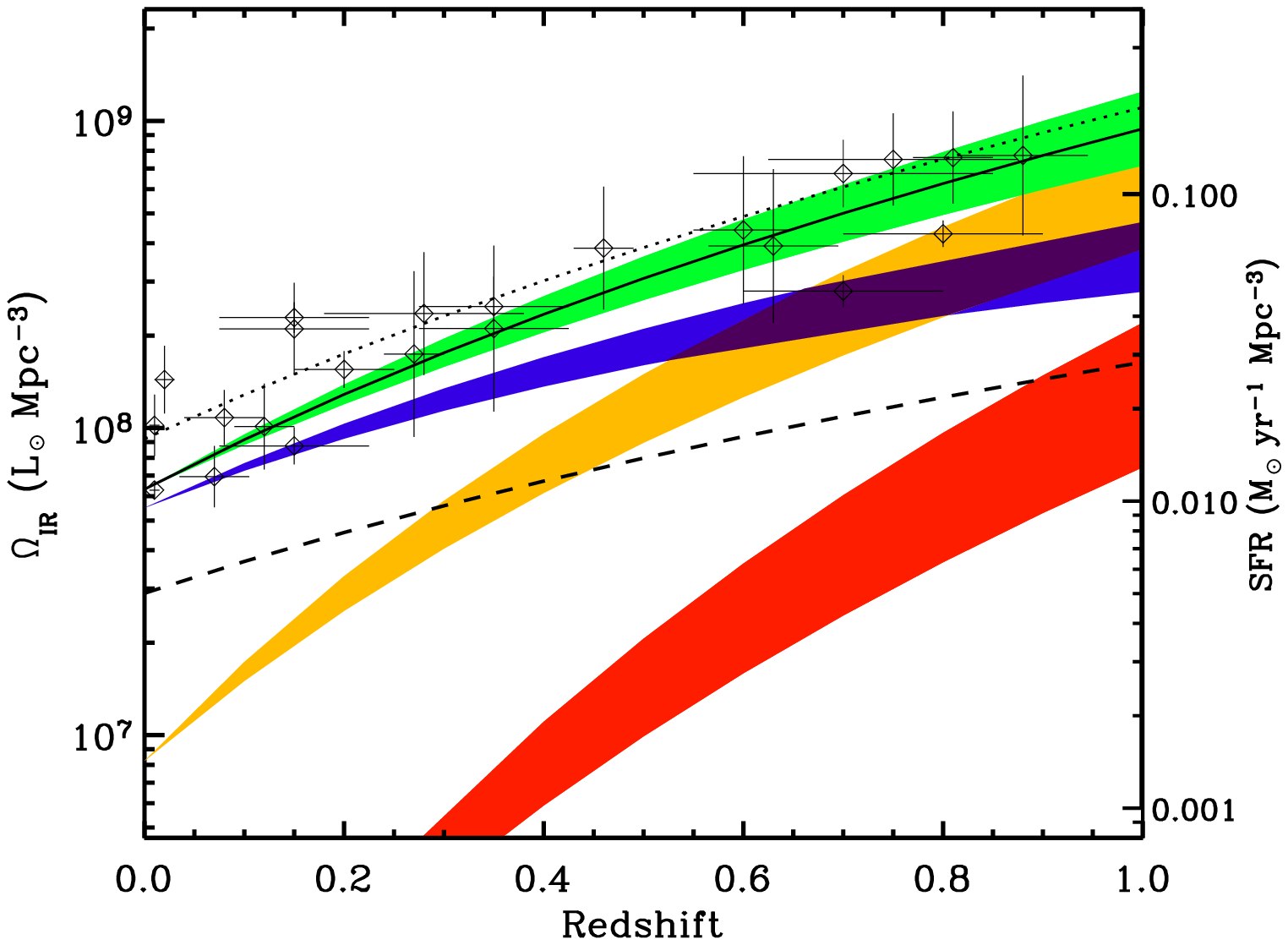}
\vskip 0.2cm \figcaption{Evolution of the comoving IR energy density
up to $z$\,=\,1 (green-filled region) and the respective contributions from
low luminosity galaxies
(i.e., L$_{\rm IR}$\,$<$\,10$^{11}$\,L$_\odot$, blue-filled
area), ``infrared luminous'' sources (i.e., L$_{\rm
IR}$\,$\geq$\,10$^{11}$\,L$_\odot$, orange-filled region) and ULIRGs
(i.e., L$_{\rm IR}$\,$\geq$\,10$^{12}$\,L$_\odot$, red-filled
region).  The lower and upper curves delimiting these regions result
from the degeneracy in the evolution of $\psi_{\rm IR}$ (see
Sect.\,7 for more details). The solid line   evolves  as (1+$z$)$^{3.9}$
and represents the best fit of the total IR luminosity
density at 0\,$\ltapp$\,$z$\,$\ltapp$\,1.
Estimates are translated into an 
``IR-equivalent SFR''
density given on the right vertical axis, where an absolute additional
uncertainty of $\sim$\,0.3\,dex should be added to reflect the dispersion
in the conversion between luminosities and SFR.
Note that the
percentage of the contribution from each population is likely independent
of this conversion.
The dashed line corresponds to the SFR measured from the UV
  luminosity {\it not corrected from dust extinction}. The dotted line
  represents the best estimate of the total star formation rate
  density as the sum of this uncorrected UV contribution and the best
  fit of the IR-SFR (solid line). At $z$\,$\sim$\,1 IR luminous
galaxies represent 70$\pm$15\% of the comoving IR energy density and
dominate the star formation activity.  Open diamonds, vertical and
horizontal bars represent integrated star formation rate densities and
their uncertainties estimated within various redshift bins and taken
from the literature 
  (\citealt{Connolly97,Tresse98,Treyer98,Flores99,Cowie99,Haarsma00,Machalski00,Sullivan01,Condon02,Sadler02,Serjeant02,Tresse02,Wilson02,Perez-Gonzalez03,Pozzi04};
  see the compilation by \citealt{Hopkins04}). 
}
\vskip .2cm
\label{fig:z_distrib}
\end{figure*}

This evolution is also represented in terms of an
``IR-equivalent SFR'' using the calibration from \citet{Kennicutt98}.
For comparison we  show  integrated star formation
rate densities estimated in various redshift bins and taken from the
literature (see the compilation by \citealt{Hopkins04} for references).
It should be noted that we have  not estimated the
contribution of the AGN IR emission  to  $\Omega_{\rm IR}$.
Following the arguments discussed in Sect.\,5.2, we believe that such
contribution 
results in a 10--15\% overestimate in the true SFR. 
On the other hand  this IR-SFR estimate does not take into account
the contribution of the unabsorbed UV light produced by the young stars.
As a result we are likely underestimating the total star formation
rate density by a factor ranging between $\sim$\,20--30\% at
$z$\,$\sim$\,1 (where dusty galaxies dominate the SFR) and
$\sim$\,60--70\% at $z$\,$\sim$\,0 (where the star-forming activity
occurs within fainter sources with low extinction). This effect
  can be seen by considering, in addition to the IR emission, the evolution of the UV
  luminosity (uncorrected for dust extinction) with redshift. Using
  GALEX data, \citet{Schiminovich05} found that the energy density
  measured at 1\,500\AA ~ evolves as (1+$z$)$^{\sim 2.5}$ at
  0\,$\ltapp$\,$z$\,$\ltapp$\,1, which is also consistent with
  previous UV measurements \citep[e.g.,][]{Lilly96}. An estimate of
  the total star formation rate density (represented as a dotted line
  in Fig.\,14) can thus be obtained by adding the contribution of the
  equivalent UV-uncorrected SFR (dashed line in Fig.\,14) to our
  best fit of the evolution of the IR star formation estimate
(converted from the energy density $\Omega_{\rm IR}$ evolving as
  (1+$z$)$^{3.9}$).
  This clearly shows that the IR emission provides
 a good approximation of the total SFR
 density at $z$\,$\sim$\,1 given the uncertainties
  affecting its current measurements (i.e., conversion between flux and
  luminosity, cosmic variance).  

One can see that the global
evolution of the luminosity function leads to a source density increase 
for both populations of IR-luminous and  low luminosity galaxies.
 However the larger increase of $L_{\rm IR}^\star$ compared to
$\phi_{\rm IR}^\star$ results in a much more rapid evolution of
sources characterized by the highest bolometric luminosities. The contribution of
IR luminous galaxies, though  negligible in the local Universe,
becomes comparable to that of normal starbursts around $z$\,$\sim$\,0.7
and  they dominate beyond.
We  also see that such IR luminous sources are mostly dominated
by LIRG-type objects up to $z$\,$\sim$\,1 and that ULIRGs still have a
modest impact ($\sim$\,10\%) at this redshift. This is once again in
good agreement with the ISOCAM surveys \citep[e.g.,][]{Aussel99}.
However these ULIRGs
have undergone the fastest evolution in the last 8\,Gyrs and our
results suggest that their contribution is likely still rising at
$z$\,$\gtapp$\,1.  As already revealed by the submillimeter surveys
\citep[see][ for a review]{Blain02}, they may  therefore  be responsible
for a very significant fraction of the star-forming activity of the
Universe at $z$\,$\sim$2--3 \citep[see also][]{Cowie04}.

How can this strong evolution observed in the infrared
be understood in the  general context of
the growth of structures~?
There is now increasing evidence that the most massive
galaxies seem to have formed their stars early in 
cosmic history and  that their
contribution 
  to the comoving SFR density of the Universe was significantly
larger at higher redshifts 
\citep[e.g.,][]{Cowie96,Cimatti04,Daddi04,Juneau05}.
 Similarly it has been shown that a significant fraction of galaxies
with masses $\mathcal{M}$\,$\gtapp$\,10$^{10}$\,$\mathcal{M}_{\odot}$
at $z$\,$\sim$\,0.7-1.0 are experiencing a violent episode of star
formation, while nearly all equally-massive sources at the present day
are found in a very quiescent mode
\citep[e.g.,][]{Flores99,Franceschini03,Bell05,Hammer05}.  In parallel
to this downsizing effect characteristic of the stellar mass assembly
history, the evolution of the luminosity function at infrared and
submillimeter wavelengths reveals
a transfer of the star formation from the most luminous high redshift
sources to more modest local starbursts. If we consider that 
LIRGs and ULIRGs
are preferentially associated with more massive systems (see
Fig.\,10d), this result might be closely
related to the mass growth picture described above.
 However the connection between the two is likely more
subtle as we found that the fraction of IR luminous sources with
$\mathcal{M}$\,$\ltapp$\,10$^{10}$\,$\mathcal{M}_{\odot}$ is not
negligible either. The relation between IR luminosities and masses has
indeed a large dispersion and the dominant contribution of LIRGs/ULIRGs
to the star formation at $z$\,$\sim$\,1 does not necessarily imply
that the SFR at these redshifts is only locked in massive systems.
A similar conclusion has been  proposed by \citet{Juneau05},
 who showed that the star formation
density at $z$\,$\sim$\,1 was dominated by systems
with 
$\mathcal{M}$\,$\ltapp$\,10$^{10.5}$\,$\mathcal{M}_{\odot}$.
Much larger samples of sources will  be required to address this
issue in more detail, especially by
 de-projecting the comoving SFR density
into the IR-luminosity/mass/redshift 3-dimensional
space. This goal could be achieved in the very near future by 
combining all the MIPS
cosmological  surveys that are currently being carried out with
{\it Spitzer.}

\section{Conclusion}

We have analyzed a sample of MIPS/{\it Spitzer} 24\mic sources
detected in the Chandra Deep Field South using ancillary optical data
from the literature (i.e., magnitudes, spectroscopic and photometric
redshifts). Our results can be summarized as follows:

\begin{itemize}
\item{} For 24\mic sources brighter than $\sim$\,80\muJypa, 
the selection criterion
 $R$\,$\ltapp$\,24\,mag provides a complete sample of
optical counterpart identifications up to $z$\,$\sim$\,0.8. 
\vspace{-.10in}
\item{} About 55-60\% of the 24\mic objects  brighter than $\sim$\,80\muJy
are located at $z$\,$\ltapp$\,1, which points to a significant
fraction of the MIPS 24\mic sources being luminous and ultra-luminous
infrared galaxies at even higher redshifts.
\vspace{-.10in}
\item{} The 24\mic source population at
0.5\,$\ltapp$\,$z$\,$\ltapp$\,1.0 is dominated by LIRG-type objects
and slightly fainter sources (i.e., L$_{\rm
IR}$\,$\ltapp$\,10$^{11}$\,L$_{\rm IR}$). ULIRGs are found to be 
rare at these redshifts.
\vspace{-.10in}
\item{} The conversion between fluxes and luminosities 
depends  strongly on the assumed infrared SEDs.
This points to
 a crucial need for more accurate determinations of
IR templates, a goal which could be achieved with MIPS and IRS.
\vspace{-.10in}
\item{} As in the local Universe, 
infrared luminous sources up to $z$\,$\sim$\,1 are also luminous
at optical wavelengths and they tend
to be more massive than the bulk of optically-selected
distant galaxies.
\vspace{-.10in}
\item{} The comoving  energy density measured in the infrared  evolves as 
(1+$z$)$^{3.9\pm0.4}$ at 0\,$\ltapp$\,$z$\,$\ltapp$\,1. In contrast, the
luminosity density in the UV only evolves as (1+$z$)$^{\sim2.5}$
over the similar redshift range. This points to a more important
extinction by dust reprocessing light in the IR at high redshift.
\vspace{-.10in}
\item{} The  infrared-selected sources at
 0\,$\ltapp$\,$z$\,$\ltapp$\,1 have
 undergone  strong evolution
characterized by $L^\star_{\rm IR}$\,$\propto$\,(1+$z$)$^{3.2_{-0.2}^{+0.7}}$
and $\phi^\star_{\rm IR}$\,$\propto$\,(1+$z$)$^{0.7_{-0.6}^{+0.2}}$. 
At $z$\,$\sim$\,1, infrared luminous galaxies (i.e., L$_{\rm
IR}$\,$\gtapp$\,10$^{11}$\,L$_\odot$) appear to be responsible for 70$\pm$15\%
of the comoving IR energy density. They dominate the
star-forming activity beyond $z$\,$\sim$\,0.7.

\end{itemize}

\acknowledgments We thank the funding from the MIPS project which is
 supported by NASA through the Jet Propulsion Laboratory (subcontract
 \#960785), as well as the {\it Spitzer\,}
Science Center for efficient technical support. 
We also appreciated the use of data products from the {\it
 Two Micron All Sky Survey\,}, which is a joint project of the
 University of Massachusetts and the Infrared Processing and Analysis
 Center (California Institute of Technology), funded by NASA and the
 National Science Foundation.  We acknowledge various worldwide teams
 for publicly providing their redshift catalogs in the Chandra Deep
 Field South, and we are particularly grateful to Jim Cadien for
 helping us in the data anlysis process. We are also indebted to
 Pierre Chanial, Ranga-Ram Chary, Daniel Dale,
David Elbaz, Carlotta
 Gruppioni, Chris Pearson and Francesca Pozzi for providing us
 with their library of IR galaxy templates and/or detailed predictions
 from their models, and we thank our referee Stephen Serjeant
for critical comments on the manuscript.
ELF is grateful to Lee Armus, Vassilis Charmandaris, Arjun Dey,
 Daniel Eisenstein, David Elbaz, George Helou, Terry Herter and Andrew
 Hopkins for stimulating discussions related to the work presented in
 this paper.

\appendix{}

\section*{Selection effects in the sample of 24\mic sources}

In this work we imposed a selection criterion
$R$\,$\leq$\,24\,mag when considering the 24\mic sources identified
with a photometric redshift from the COMBO-17 survey.  As we showed in
Sect.\,3, this selection obviously results in an incompleteness of
the MIPS-selected population at $z$\,$\gtapp$\,0.8-1.0. 
We provide in this Appendix a few more details characterizing
the possible nature of the sources that we may have missed
in the estimate of the LFs at high redshift.

In Figure\,15 we show the relation between the $R-24\mu m$ color and
the $R$-band magnitude of the MIPS sources identified with an optical
counterpart from COMBO-17 (top panel). The 24\mic Vega magnitudes were calculated
assuming a Zero Point of 7.3\,Jy as defined in the {\it Spitzer
Observing Manual.}\footnote{An electronic version is available at
http://ssc.spitzer.caltech.edu/documents/SOM/} 
As pointed out in Sect.\,6, the apparently well-defined
correlation that is observed is a natural
consequence of the small range of 24\mic fluxes (typically a factor of
$\sim$\,10 over our sample) compared to the much larger range
($\sim$\,5 magnitudes) covered by the optical counterparts of the MIPS
sources (see also Fig.\,8a).  If we only consider 24\mic objects
brighter than the 80\% completeness of our survey (i.e., $f_{24\mu
m}$\,$\geq$\,83\muJypa), we see that the selection at $R$\,=\,24\,mag
defines a complete sample up to  $R-24\mu m$\,$\sim$\,11.6. We miss
therefore the reddest sources  of the MIPS catalog (shaded region
of Fig.\,1).

\begin{figure*}[htpb]
\begin{center}
\includegraphics*[width=15cm]{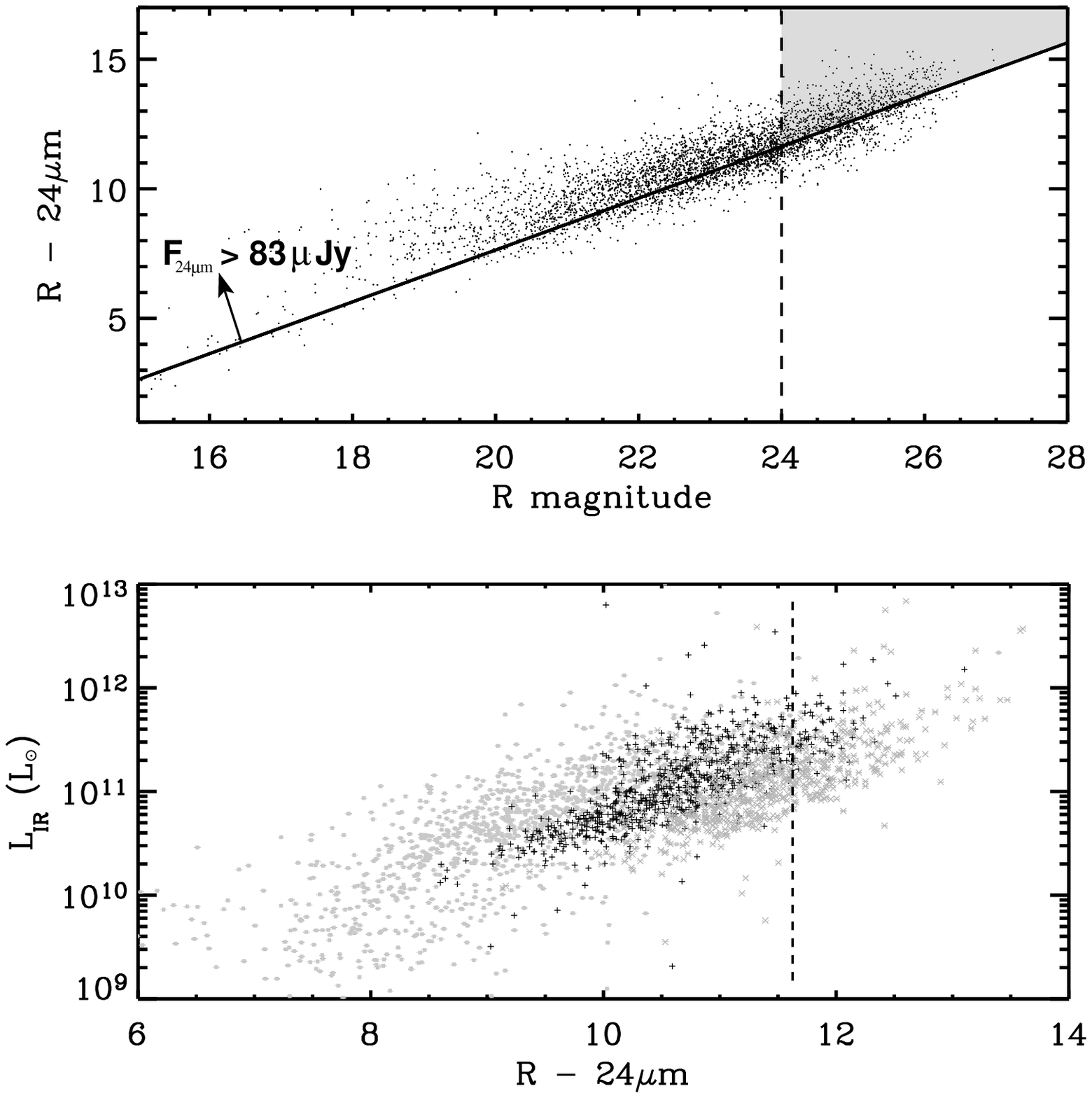}
\vskip 0.2cm \figcaption{{\it Top:\,} The $R$\,--\,24\mic color as a
function of the $R$-band magnitude for the MIPS sources identified
with an optical counterpart in COMBO-17. Sources above the solid line
are brighter than the 80\% completeness limit of the MIPS 24\mic
survey (i.e., $f_{24\mu m}$\,$\geq$\,83\muJypa). Given the selection
criterion $R$\,$\leq$\,24\,mag (dashed line), sources redder than
$R$\,--\,24\micpa\,$\sim$\,11.6 in the shaded region are therefore 
missed when deriving the luminosity functions in Sect.\,7.  {\it
Bottom:\,} The total IR luminosities (see Sect.\,5) versus the
$R$\,--\,24\mic color for the MIPS sources with $R$\,$\leq$\,22\,mag
(grey filled squares), 22\,mag\,$<$\,$R$\,$\leq$\,23\,mag ('$+$' black
symbols) and 23\,mag\,$<$\,$R$\,$\leq$\,24\,mag ('$\times$'
grey symbols). The vertical dashed line represents the color selection
 $R$\,--\,24\micpa\,$=$\,11.6.  Note that the reddest sources
still span a wide range of IR luminosities.  }
\vskip .2cm
\end{center}
\end{figure*}

To get a hint into the possible nature of these targets, we also
plot in figure\,15 the total infrared luminosities (as derived in
Sect.\,5) versus the $R-24\mu m$ colors for the MIPS sources
identified with a spectroscopic or a photometric redshift (bottom panel). Different
symbols are used to highlight sources as a function of the $R$-band
magnitude.  As we see, 24\mic objects redder than $R-24\mu
m$\,$\sim$\,11.6 are not obviously the most extreme in terms of dust
emission and their luminosity can vary between
$\sim$\,5$\times$10$^{10}$\,L$_\odot$ and a few~10$^{12}$\,L$_\odot$ in the
infrared.  Even though sources fainter than $R$\,$\sim$\,24\,mag may
follow a different trend, this dispersion of L$_{\rm IR}$ with both the
color and the $R$-band magnitude suggests that these sources 
span a wide range of IR luminosities. Taking into account the evolution of the sensivity limit
with redshift
in our 24\mic survey (see Fig.\,7), they should thus  be located in a wide range of redshifts 
at $z$\,$\gtapp$\,1.

\end{document}